\documentclass[%
reprint
]{revtex4-2}

\usepackage{color}
\usepackage[cmex10]{amsmath}
\usepackage{amssymb}
\usepackage[dvipsnames]{xcolor}
\usepackage{graphicx}
\usepackage{subcaption}
\usepackage{booktabs}
\usepackage{bm}
\usepackage{subfiles}
\usepackage{multirow}
\usepackage{amsmath}
\usepackage{mathtools}
\usepackage{listings}

\mathchardef\mhyphen="2D

\usepackage{graphicx}
\usepackage{dcolumn}
\usepackage{bm}

\begin{document}

\title{Analytical Scattering Solution of Huygens' Refracting Metasurfaces Reveals Unknown Functionalities}

\author{Gleb A. Egorov}
\author{George V. Eleftheriades}%
\affiliation{
	Dept. of Electrical and Computer Engineering, University of Toronto, Toronto, Canada
}

\date{\today}

\begin{abstract}
	Scattering of refracting Huygens' metasurfaces is revisited. A new analytical closed-form solution is obtained for the two-dimensional problem under transverse electric (TE) plane-wave incidence -- a solution which gives the scattering for any (in-plane) angle of incidence and frequency. In the process of obtaining the solution, interesting mathematical concepts are considered -- such as solving and enforcing uniqueness upon an infinite system of equations. The obtained solution uncovers two previously unknown functionalities of refracting Huygens' metasurfaces -- namely reflectionless refraction and oblique absorption.
\end{abstract}

\maketitle


\section{Introduction}

Laws of refraction between different media are well-known and celebrated -- Snell's laws dictate the directions of refracted/reflected waves while Fresnel equations quantify the amplitudes of these waves. Together, they completely specify the behavior of electromagnetic waves in the presence of a material discontinuity \cite{jackson1999classical,pozar2011microwave,ishimaru2017electromagnetic}. In the context of this standard theory, the question of refraction/reflection within a single homogeneous medium is a meaningless one. Recent work (relative to the laws just mentioned) of Selvanayagam \cite{selvanayagam2013discontinuous}, Pfeiffer \cite{pfeiffer2013metamaterial} and Epstein \cite{epstein2014floquet} investigated refraction via a thin (with respect to the wavelength of interest) boundary between two half-spaces of identical media. Selvanayagam found that a boundary possessing electric and magnetic responses was indeed capable of achieving such refraction, where electric/magnetic response means that a tangential electric/magnetic field on the surface induces electric/magnetic dipole moments respectively. The required electric and magnetic dipole moments were found to be in-plane of the boundary and orthogonal to each other -- subsequently, such a boundary was termed a Huygens' metasurface. Furthermore, the surface parameters specifying the coupling strength between the acting fields and induced dipoles varied along the metasurface in a periodic fashion, which is reminiscent of conventional diffraction gratings. Indeed, the orthogonal electric and magnetic responses with the correct spatial dependence constituted an extremely well-designed diffraction grating -- one which would not excite any diffraction orders other than the refracted and reflected waves when illuminated at the designed/desired incidence angle and frequency. 

In theoretical terms of \cite{selvanayagam2013discontinuous,pfeiffer2013metamaterial,epstein2014floquet}, these metasurfaces were treated as sheets of continuously-varying surface electric impedance $Z$ and surface magnetic admittance $Y$, and together comprising an overall $ZY$ boundary. In this formalism, the oscillating electric/magnetic dipole moments were treated as electric/magnetic surface currents. Via this viewpoint Selvanayagam et al. was able to show that in order for the surface to refract (without producing undesired diffraction orders), the $Z$ and $Y$ both must have a cotangent variation along the boundary (with an appropriate amplitude scaling for both). Despite this significant progress regarding the theory of refraction, a question that remained unanswered until now is how will this postulated mathematical boundary scatter when the incidence upon it is at some non-designed angle and/or frequency.

In this letter, the TE scattering of the $ZY$ refracting metasurface is studied and solved analytically. Apart from providing the closed-form expressions, which are of interest in themselves, two new functionalities of $ZY$ boundaries are discovered -- reflectionless refraction and oblique absorption. 

\section{The Scattering Solution}

\subsection{Conventions and Definitions} \label{sec:ConvDefs}

\begin{figure}
	\centering
	\includegraphics[width=0.4\textwidth]{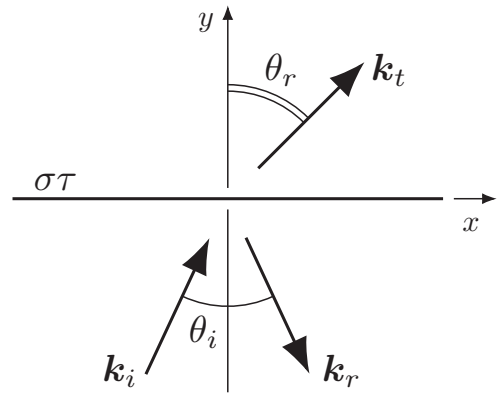}
	\caption{A refracting $\sigma\tau$ boundary, subject to designed incidence.}
	\label{fig:IdealRefractionSigmaTau}
\end{figure}

The problem is time-harmonic, with the assumed $e^{j \omega' t}$ time dependence. The refracting boundary is placed in free space, on the $y=0$ plane (see Fig. \ref{fig:IdealRefractionSigmaTau}). Surface parameters are functions of the $x$-coordinate only. The incident waves which we will consider are TE polarized ($E_z \neq 0$, while $E_x=E_y=0$), do not propagate along the $z$-direction (incident wavevectors have zero $z$-component) and are incident upon the refracting boundary from below (incident wavevectors have positive $y$-components). These choices, in part, lead to the problem being two-dimensional -- the incident/scattered fields and induced currents are independent of the $z$-coordinate.

In the context of this letter, if a surface refracts then it behaves as is shown in Fig. \ref{fig:IdealRefractionSigmaTau}. Refraction is said to occur when only three waves are present -- the incident wave from below, a single refracted wave above, and a single reflected wave below the refracting boundary. If other diffraction gratings are present, the boundary is said to diffract/scatter.

Instead of using surface electric impedances and magnetic admittances ($Z$ and $Y$) we choose surface electric and magnetic conductivities ($\sigma$ and $\tau$), defined via Ohm-like relations
\begin{subequations}
\begin{align}
	J_z(x,y=0) &= \sigma(x) \frac{E_z(x,0^+)+E_z(x,0^-)}{2}, \\
	M_x(x,y=0) &= \tau(x) \frac{H_x(x,0^+)+H_x(x,0^-)}{2},
\end{align}
\end{subequations}
where $J_z/M_x$ are the induced $z/x$-directed surface electric/magnetic current densities (units of current per length), $E_z/H_x$ are the total (incident plus scattered) tangential electric/magnetic fields acting on the surface, and $E_z(x,0^\pm)=\lim_{\epsilon\to0}E_z(x,\pm|\epsilon|)$ (and likewise for $H_x$). See supplementary material \cite{supp} for why the averaging of the fields is required and \cite{kuester2003averaged} for a discussion on the validity of this averaging for physical structures. The reason for using $\sigma$ and $\tau$ instead of $Z$ and $Y$ is because the former appear more naturally in Maxwell's equations. 

We know from the work in \cite{selvanayagam2013discontinuous} that a refracting metasurface located in free space and ``designed" for refraction from the incident angle $\theta_i$ towards the refracted angle $\theta_r$ (with respect to the surface normal) at frequency $\omega$ is given by (while accounting for the differences in conventions)
\begin{subequations} \label{eq:LosslessSurf}
	\begin{align}
		\sigma(x) &= j \frac{2k_1}{\eta k} \tan \left(\frac{k_s x}{2}\right),\\
		\tau(x) &= j \frac{2\eta k}{k_1} \tan\left(\frac{k_s x}{2} \right),
	\end{align}
\end{subequations}
where $k = \omega/c$ ($c$ being the speed of light in vacuum), $\eta$ is the impedance of free space, $k_1 = k \cos\theta_r$ and $k_s = k(\sin\theta_r-\sin\theta_i)$. The geometry of such a boundary subject to designed incidence is shown in Fig. \ref{fig:IdealRefractionSigmaTau}. Note that these surface conductivities are purely imaginary -- Poynting's theorem can be used to deduce that such a surface is passive and lossless. Passivity and losslessness are important and desired properties -- the former implies that the surface does not need any internal energy sources to facilitate the refraction, while the latter means no incident energy is wasted via heat.

Because the $\sigma\tau$ boundary is periodic, Floquet scattering applies (see supplementary material \cite{supp}), which dictates what form the scattered fields can take given some particular incident wave. For an incident wave of frequency $\omega'$ (and wavenumber $k'=\omega'/c$) and of the form
\begin{equation}
	\mathbf{E}_i = \hat{\bm z} E_i e^{-j(k_x' x + k_y' y)},
\end{equation}
where $E_i$ is a complex amplitude, Floquet scattering dictates that the most general form of the total (incident plus scattered) electric field for our two-dimensional problem can be written as
\begin{subequations}
\begin{align}
	E_z(x,y>0) &= E_i \sum_{a=-\infty}^\infty \mathrm{T}_a e^{-j(k_x'+k_s a) x - j k_a' y},\\
	E_z(x,y<0) &= E_i \bigg(e^{-jk'_x x - jk_0' y} \hspace{5pt} +  \nonumber \\ & \sum_{a=-\infty}^\infty \Gamma_a e^{-j(k_x'+k_s a) x + j k_a' y}\bigg),
\end{align}	
\end{subequations}
where $\mathrm{T}_a/\Gamma_a$ are the transmission/reflection coefficients for the $a^\text{th}$ Floquet mode, and $k'_a = \pm\sqrt{k'^2-(k_x'+k_s a)^2}$ with the plus sign when the square root is real and minus otherwise (again, see supplementary material \cite{supp}). 

Solving the scattering problem entails obtaining expressions for the aforementioned transmission and reflection coefficients. Once known, the total electric field is known, and the total magnetic field is easily obtained from the standard plane-wave relations of electromagnetics.

\subsection{Outline of Solution and T/$\mathbf{\Gamma}$ Expressions}

Let us now provide an outline of how to obtain the transmission/reflection coefficients. The tedious mathematical reasoning is provided in full detail in the supplementary material \cite{supp}.

A surface electric/magnetic current creates a discontinuity in magnetic/electric fields, respectively, which is quantified via the well-known boundary conditions of electromagnetics \cite{harrington2001harrington}. Using these conditions together with the definitions of Sec. \ref{sec:ConvDefs} allows one to write a system of linear equations for the scattered mode amplitudes, which happens to be infinite due to the infinite number of unknown Floquet modes. The system of equations has a somewhat simple form, and all possible solutions to it can be easily written in terms of a single complex coefficient (which can be picked at will). Note that, unlike for finite systems of equations, a non-degenerate infinite system can have infinitely many mathematically valid solutions. 

This abundance of valid solutions is undesired from a physical perspective -- as intuition dictates, a physical boundary subject to a particular incident wave must have a unique solution. Studying the valid solutions under the incidence case $k'=k$ and $k_x'=k_x$ (i.e. the incidence being the one for which the boundary was designed) one indeed can argue that all of the possible solutions are unphysical except one (the total fields of the solutions, except one, do not converge on the surface). Placing the constraint that the obtained fields must converge everywhere leads us to the single, unique solution, in agreement with the previous work of \cite{selvanayagam2013discontinuous}. However, the same cannot be said for other incidence scenarios ($k'\neq k$ and/or $k_x'\neq k_x$) -- there, the fields of every possible solution exhibit non-converging behavior. This implies that a boundary given by (\ref{eq:LosslessSurf}) is ill-defined in the sense that no physically valid scattering solutions exist (for non-designed incidence). 

We remedy the situation by redefining the refracting boundary to include a small (but finite) amount of loss, via two independent loss mechanisms $\Delta$ and $\Lambda$, as 
\begin{subequations}
	\begin{align}
		\sigma_{\Delta\Lambda}(x) &= j\frac{2k_1}{\eta k}\frac{\sin(k_s x/2)}{\cos(k_s x/2)+j\Delta_\sigma\sin(k_s x/2)}+\frac{2}{\eta k}\Lambda_\sigma, \\
		\tau_{\Delta\Lambda}(x) &= j\frac{2\eta k}{k_1}\frac{\sin(k_s x/2)}{\cos(k_s x/2)+j\Delta_\tau\sin(k_s x/2)}+\frac{2\eta}{k_1}\Lambda_\tau,
	\end{align}
\end{subequations}
where $\Delta/\Lambda$ are small positive numbers, and whose subscripts (as seen in the expressions above) signify that the electric and magnetic losses are independent. One can easily check via Poynting's theorem that indeed these terms signify loss, and not gain (as is expected for a passive metasurface). Note that these surface parameters are still tangent-like, except the infinities and the zeros of the tangents have been eliminated. Again, an infinite linear system of equations can be written, but now uniqueness can be imposed for all incidence scenarios (any $k'$, $k'_x$) -- only one solution will lead to converging fields on the surface. The expressions for the Floquet mode amplitudes for this slightly lossy refracting surface are quite cumbersome and are given in the supplementary material \cite{supp}, while their limit of zero loss ($\Delta,\Lambda \to 0$) exists and is given by
\begin{subequations} \label{eq:Solution}
	\begin{align}
		\mathrm{T}_{a<1} &= 0, \\
		\mathrm{T}_1 &= \frac{4k_1 k'_0 \frac{k}{k'}}{\left(k_1 +\frac{k}{k'} k'_1\right)\left(k_1+\frac{k}{k'}k'_0\right)}, \\
		\mathrm{T}_{a>1} &= \frac{(-1)^{a+1}+1}{k_1+\frac{k}{k'} k_a'}\cdot\frac{2k_1 k'_0 \frac{k}{k'}}{k_1+\frac{k}{k'} k'_0} \prod_{n=1}^{a-1}\frac{\frac{k}{k'} k'_n-k_1}{\frac{k}{k'}k'_n + k_1},\\
		\Gamma_{a<0} &=0, \\
		\Gamma_0 &= -\frac{k_1-\frac{k}{k'} k'_0}{k_1+\frac{k}{k'} k'_0}, \\
		\Gamma_1 &= 0, \\
		\Gamma_{a>1} &= \frac{(-1)^{a+1}-1}{k_1+\frac{k}{k'} k_a'}\cdot\frac{2k_1 k'_0 \frac{k}{k'}}{k_1+\frac{k}{k'} k'_0} \prod_{n=1}^{a-1}\frac{\frac{k}{k'} k'_n-k_1}{\frac{k}{k'}k'_n + k_1}.
	\end{align} 
\end{subequations}
These equations, obtained as the limit of zero loss, provide the asymptotic behavior of physical low-loss refracting 'tangent-like' $\sigma\tau$ surfaces.

\subsection{New Refraction Effects and Other Properties}

Looking at the numerators of (\ref{eq:Solution}), especially the $(-1)^{a+1}\pm 1$ of (\ref{eq:Solution}c) and (\ref{eq:Solution}g), we see that $\Gamma_\text{odd}=\mathrm{T}_{\text{even}}=0$. 

Note that $\Gamma_0$ is zero when $k'_0/k'=k_1/k$. This condition can be re-written as $\theta'_i=\theta_r$ -- specular reflection will be zero when the angle of incidence ($\theta'_i$) is equal to the designed refraction angle ($\theta_r$), which remains true at any frequency. We now consider incidence at the design frequency by setting $k'=k$. It is possible to design a surface (by providing the values of $k, k_x, k_s$) for which $k'_1 \in \mathbb{R}$ and $k'_2 \in j\mathbb{R}$ when $k'_0=k_1$. For example, choosing $k_x=0.2k$ and $k_s=0.3k$ indeed leads to a real $k'_1$ and an imaginary $k'_2$ when $k'_0=k_1$ and $k'=k$, as one can easily check. Let us dwell on what this scenario (choice of surface and illumination angle) accomplishes -- transmitted wave is refracted with no specular reflection while all other modes are evanescent. The evanescent modes are localized to the surface and are not visible to observers far away from it. What these observers do see is a $\sigma\tau$ boundary refracting an incident plane wave with no specular reflection. Up to now it was believed that such feat could not be accomplished by a $ZY$ (i.e. $\sigma\tau$) boundary, but could only be attained by a more complicated boundary (one with magneto-electric coupling, see \cite{epstein2016arbitrary,asadchy2016metasurfaces}).

Another scenario of interest is a surface (with some given $k, k_x, k_s$) illuminated at $k'=k$ and $k'_0=k_1$ (again making $\Gamma_0=0$), but for which $k'_1 \in j\mathbb{R}$. For example, choosing $k_x=0.2k$ and $k_s=0.5k$ indeed leads to $k'_{a>0}\in j\mathbb{R}$. In this case even the first refracted mode is evanescent, and the incident wave simply gets absorbed by the surface with no reflection. And so, we find another unknown functionality of a $\sigma\tau$ boundary -- in the limit of zero loss these surfaces can act as reflection-less absorbers under oblique incidence (note that some non-zero loss is required). 

Finally, note that $\Gamma_0$ is independent of frequency. This can be seen when it is written as $\Gamma_0=-\frac{k_1/k-k'_0/k'}{k_1/k+k'_0/k'}$. Since $k_1$ is real (the surface is designed to refract) $k_1/k$ can be interpreted as the sine of the designed refraction angle (i.e. $\sin\theta_r$). Since we always excite the boundary with a propagating plane wave, $k'_0$ is also real and $k'_0/k'$ can be interpreted as the sine of the incident angle (i.e. $\sin\theta'_i$). Thus, we have $\Gamma_0=-\frac{\sin\theta_r-\sin\theta'_i}{\sin\theta_r+\sin\theta'_i}$ which is clearly independent of frequency. The same cannot be said for other non-zero $\Gamma_a$ and $\mathrm{T}_a$ coefficients.

\section{Case Studies}

Let us now investigate our analytical scattering solution graphically. Without loss of generality we set $E_i=1$ and leave $k$ variable.

\begin{figure*}[t!]
	\centering
	\begin{subfigure}{0.24\textwidth}
		\centering
		\includegraphics[width=1\textwidth]{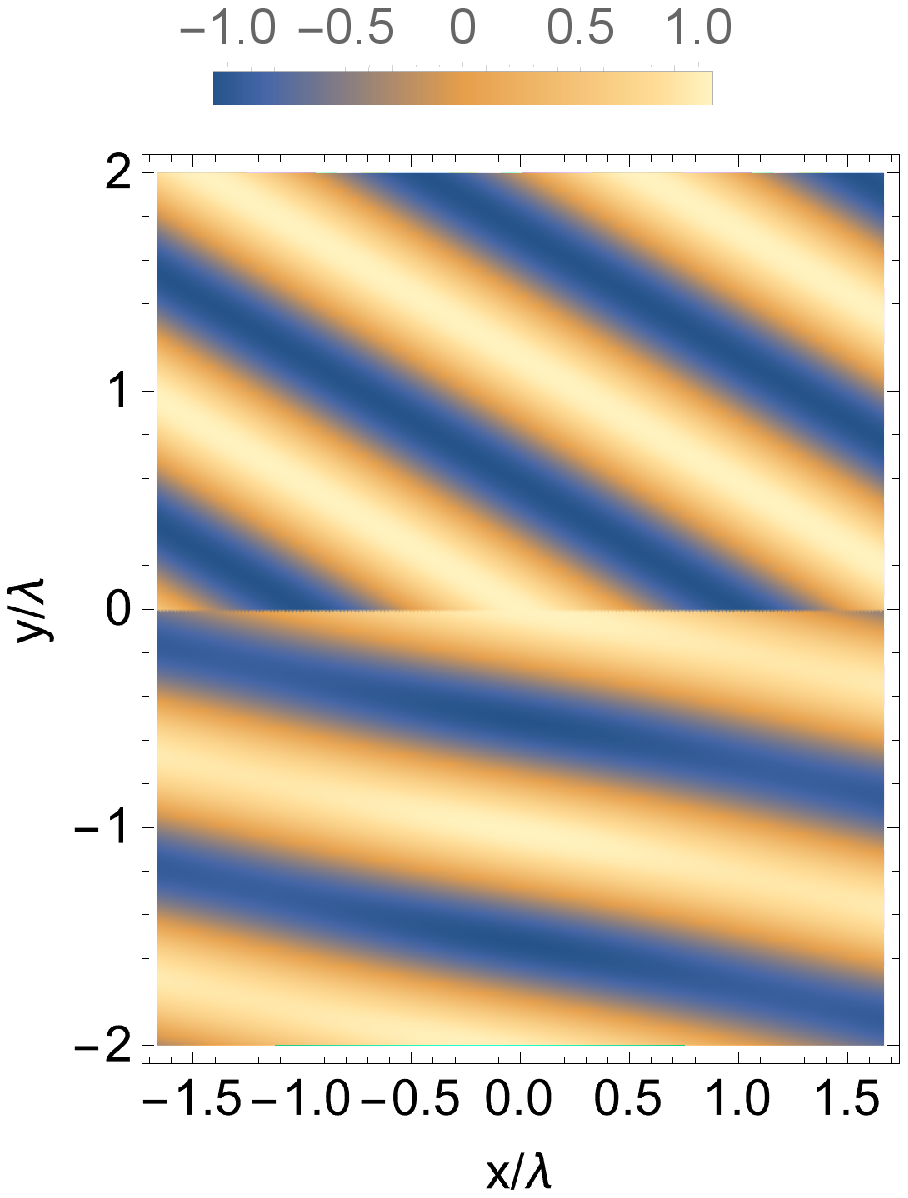}
		\caption{$k_s=0.3k$, $k_x=0.2k$, $k'=k$, $k'_x=0.2k$}
	\end{subfigure}
	\hfill
	\begin{subfigure}{0.24\textwidth}
		\centering
		\includegraphics[width=1\textwidth]{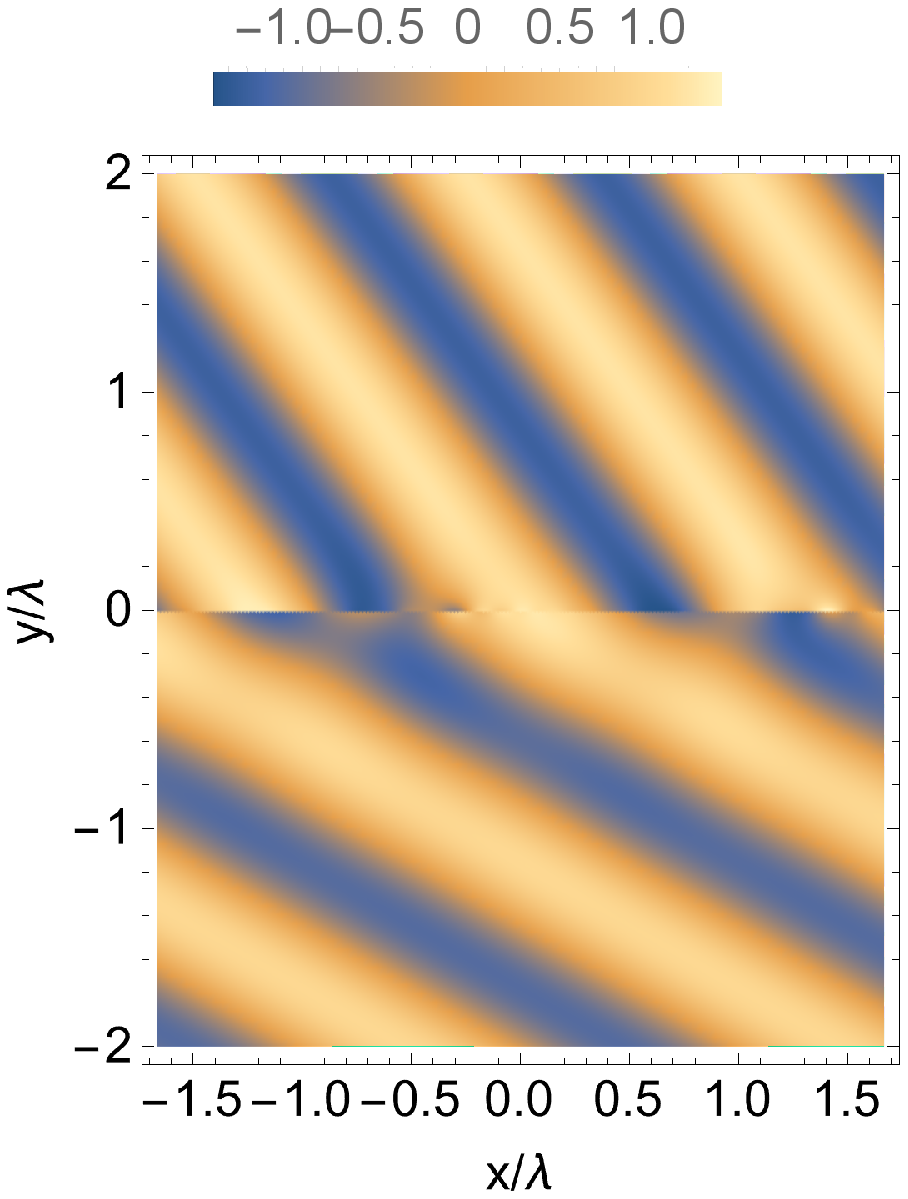}
		\caption{$k_s=0.3k$, $k_x=0.2k$, $k'=k$, $k'_x=0.5k$}
	\end{subfigure}
	\hfill
	\begin{subfigure}{0.24\textwidth}
		\centering
		\includegraphics[width=1\textwidth]{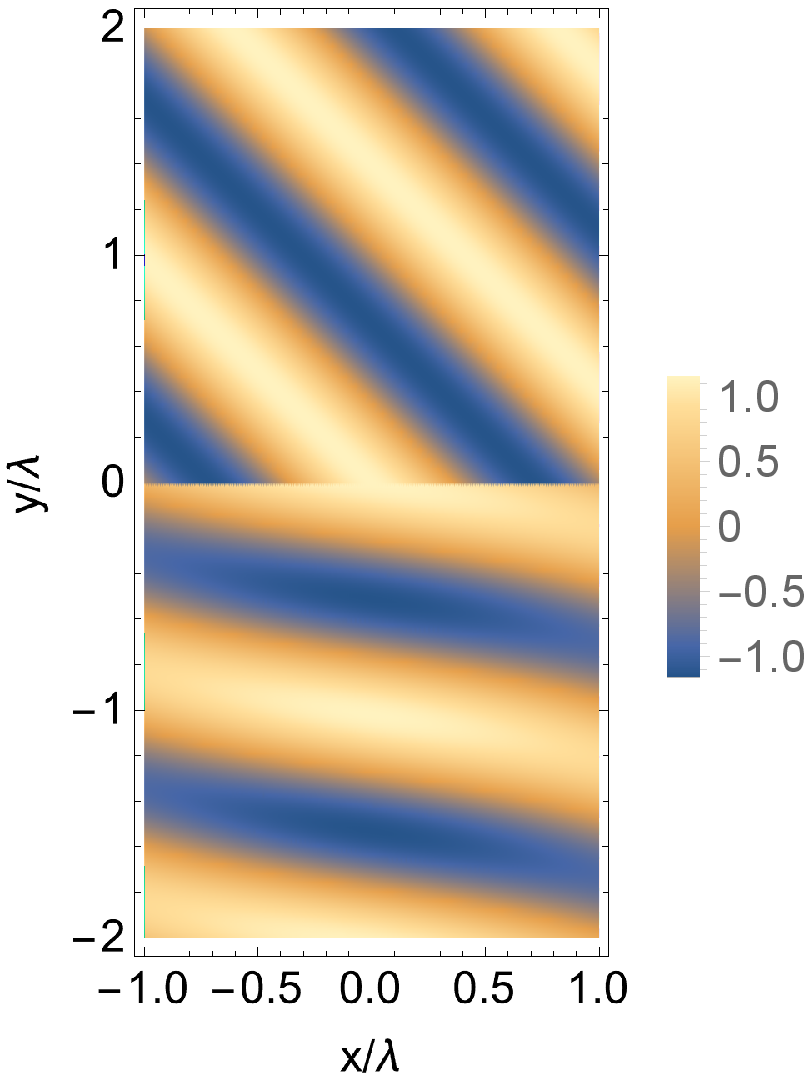}
		\caption{$k_s=0.5k$, $k_x=0.2k$, $k'=k$, $k'_x=0.2k$}
	\end{subfigure}
	\hfill
	\begin{subfigure}{0.24\textwidth}
		\centering
		\includegraphics[width=1\textwidth]{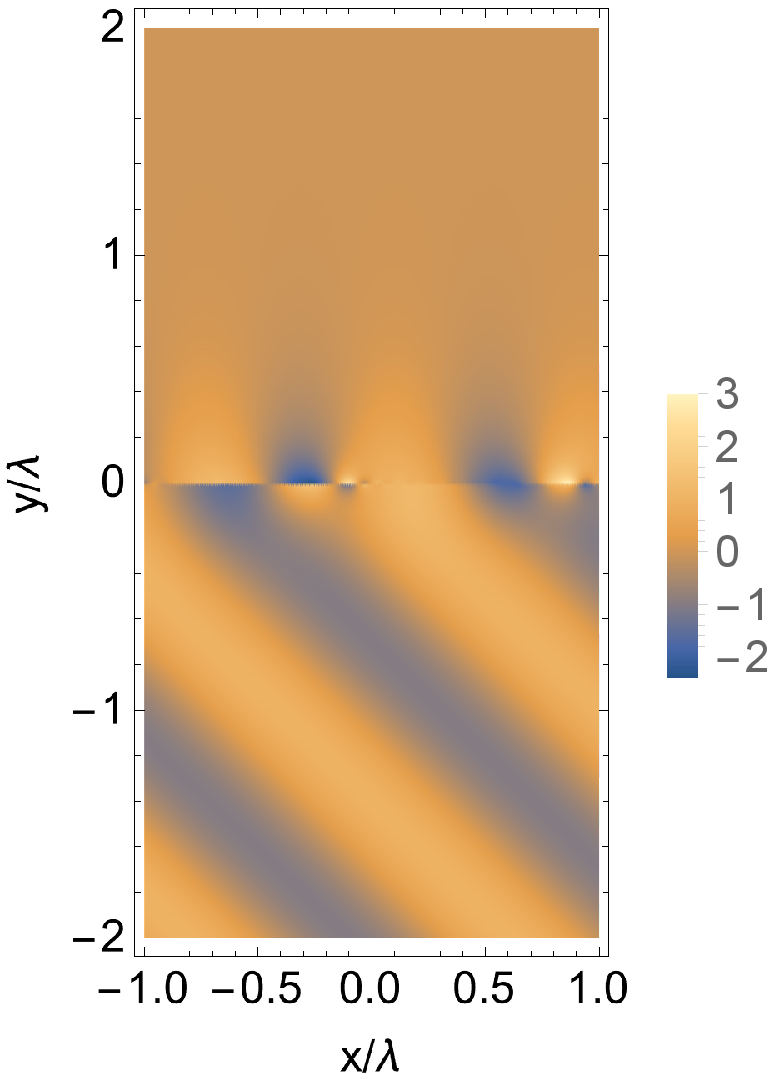}
		\caption{$k_s=0.5k$, $k_x=0.2k$, $k'=k$, $k'_x=0.7k$}
	\end{subfigure}
	\caption{$\Re\{E_z(x,y)\}$ is plotted for two $\sigma\tau$ surfaces -- (a-b) show a surface with $k_s=0.3k$ and $k_x=0.2k$, and in (c-d) the surface is given by $k_s=0.5k$ and $k_x=0.2k$. In (a) and (c) the surfaces are subject to designed incidence (in angle and frequency); in (b) and (d) the surfaces are illuminated at the design frequency but with the incident angle satisfying $\theta'_i=\theta_r$. Note that $\lambda$, which appears in the scaling of the $x$ and $y$ coordinates, is the free-space wavelength at the design frequency ($\lambda=2\pi/k$).}\label{fig:SigmaTauFields}
\end{figure*}

We first study surfaces with design parameters $k_s=0.3k$ and $k_x=0.2k$ (note that a particular choice of $k$ and $k_x$ set the design frequency and design incident angle) -- this is a surface that achieves refraction (only the incident, reflected and refracted waves exist) for the incident angle $\theta_i$ of 11.5$^\circ$ (in this case the refracted beam angle $\theta_r$ is 30$^\circ$). We plot the electric field by evaluating a partial sum of all the $\mathrm{T}_a/\Gamma_a$ waves -- we chose to use the first 20 Floquet modes ($a_{\max} =20$). Well before the 20$^\text{th}$ mode, the waves become evanescent (localized to the vicinity of the surface), and our choice of $a_{\max}$ is sufficient for graphical representation. Figure \ref{fig:SigmaTauFields}(a-b) shows the $\Re\{E_z(x,y)\}$ (i.e. electric field at the instant $t=0$) for various scenarios. In (a), the surface is subject to designed incidence (in both angle and frequency). Indeed, we see refraction -- a single uniform wavefront is seen above the surface, while below a small ripple is observed due to non-zero specular reflection. In (b), the surface is subject to incidence at the designed frequency but with the incidence angle $\theta'_i=\theta_r=30^\circ$ -- the aforementioned condition required for ``reflection-less" refraction with a $\sigma\tau$ boundary. Note that this occurs because at this incidence angle, the electric and magnetic contributions to reflection are equal in magnitude but opposite in phase and hence cancel to produce zero reflection. Indeed, below the surface a single uniform wavefront is observed with no specular reflection.

A $\sigma\tau$ with $k_s=0.5k$ and designed for $k_x=0.2k$ is analyzed in a similar manner. Figure \ref{fig:SigmaTauFields}(c-d) shows the truncated (again $a_{\max}=20$) $E_z$ for two incidence scenarios. In part (c), refraction is seen (with a stronger specular reflection than before) under designed incidence;  in (d), the aforementioned case of an oblique absorber (occurs when $\theta'_i=\theta_r=44^\circ$ and $k_1 \in j\mathbb{R}$) is shown.

More numerical evaluations are provided in the supplementary material \cite{supp}.

		
\section{Conclusion}

In this letter the scattering problem of refracting Huygens' metasurfaces was analytically solved. Previously unknown functionalities of said boundaries were uncovered from the solution.

Current work solves the scattering problem for in-plane TE incidence ($E_x=E_y=0$ and $k_z=0$). The presented method can be applied to extend the scattering solution to TM incidence and non-planar incidence.

\bibliography{ms}
	
\end{document}


\title{Supplementary Material to ``Analytical Scattering Solution of Huygens' Refracting Metasurfaces Reveals Unknown Functionalities"}
	
	\author[1]{Gleb A. Egorov}
	\author[1]{George V. Eleftheriades}
	\affil[1]{Dept. of Electrical and Computer Engineering, University of Toronto, Toronto, Canada}
	
	\date{\today}

\maketitle

\section{Background and Definitions}

\subsection{TE Fields Produced by Harmonic Current Sheets} \label{sec:HarmonicCurrents}

An important aspect of classical electromagnetic theory is the well-known boundary conditions \cite{pozar2011microwave,harrington1961time}. For metasurfaces which are flat and surrounded on both sides by free space the conditions can be written as \cite{pozar2011microwave,harrington1961time}
\begin{subequations} \label{eq:BoundaryCondBackground}
	\begin{align}
		\hat{\bm n} \times \left(\mathbf{E}_2-\mathbf{E}_1\right) &= -\mathbf{M}, \\
		\hat{\bm n} \times \left(\mathbf{H}_2-\mathbf{H}_1\right) &= \mathbf{J}, \\
		\hat{\bm n} \cdot \left(\mathbf{E}_2-\mathbf{E}_1\right) &= \frac{\rho_e}{\epsilon_0}, \\
		\hat{\bm n} \cdot \left(\mathbf{H}_2-\mathbf{H}_1\right) &= \frac{\rho_m}{\mu_0},
	\end{align}
\end{subequations}
where $\hat{\bm n}$ is a direction vector normal to the surface of the discontinuity, $\mathbf{E}/\mathbf{H}_{1/2}$ represents the electric/magnetic field immediately below/above the discontinuity (by definition, $\hat{\bm n}$ points from ``below" to ``above"), $\mathbf{M}/\mathbf{J}$ are the magnetic/electric surface current density (per unit length) whose directions are perpendicular to $\hat{\bm n}$ (currents are constrained to the surface of the discontinuity), and $\rho_{e/m}$ are the electric/magnetic surface charge densities (per unit area). The above equations are point-wise relations -- all quantities appearing (except $\hat{\bm n}$ due to the flat discontinuity case) are functions of the two spatial coordinates along the surface, and the terms are evaluated at the same location on the discontinuity \cite{pozar2011microwave,harrington1961time}. Note that while we provided all four of the standard boundary conditions, we will only use the first two -- the ones in which $\mathbf{J}/\mathbf{M}$ appear.

On many occasions we will be interested in the fields produced by harmonic surface currents. Assume the surface currents to lie in the $y=0$ plane, and are functions only of $x$ (forcing the problem to be 2D, and all fields be independent of the $z$-coordinate). Furthermore, we will be mostly interested in TE-polarized fields. According to the electromagnetic boundary conditions, the only non-zero current components which support TE fields are $J_z$ and $M_x$. And so consider a sheet of electric surface current to have the form
\begin{equation}
	J_z(x,y=0) = J_0 e^{-j k_x x}.
\end{equation}
This current creates a $z$-directed magnetic vector potential $A_z$, from which the fields can be found. From symmetry arguments, it is easy to see that $A_z(x,y) = A_z(x,-y)$. From this, it is obvious that whatever fields are created by the current, they must satisfy $|\mathbf{F}(x,y)|=|\mathbf{F}(x,-y)|$, where $\mathbf{F}$ stands for both the electric and magnetic fields. Using this observation together with boundary conditions that the current sheet must satisfy, the current must produce magnetic fields which behave as $H_x(x,y>0) = -H_x(x,y<0)$. Furthermore, since the currents are creating the fields, whatever waves exist must propagate away from the current. All these observations constrain the possible solutions of the electric current boundary condition to have a unique solution which is
\begin{subequations}
	\begin{align}
		E_z(x,y>0) &= -\frac{J_0\eta k}{2k_y}e^{-j(k_x x+k_y y)},    &    E_z(x,y<0) &= -\frac{J_0\eta k}{2k_y}e^{-j(k_x x-k_y y)}, \\
		\mathbf{H}(x,y>0) &= -\frac{J_0}{2} e^{-j(k_x x+k_y y)} 
		\begin{bmatrix}
			1 \\
			-k_x/k_y \\
			0
		\end{bmatrix}, & 
		\mathbf{H}(x,y<0) &= \frac{J_0}{2} e^{-j(k_x x-k_y y)} 
		\begin{bmatrix}
			1 \\
			k_x/k_y \\
			0
		\end{bmatrix},
	\end{align}
\end{subequations}
where $k_y = \sqrt{k^2-k_x^2}$. In the case when $k>|k_x|$, the positive choice of the square root makes the waves propagate away from the surface. In the case when $k<|k_x|$, $k_y$ will be imaginary and the surface will support evanescent waves. In order for the evanescent waves to decay away from the current (and have finite power), the negative square root has to be taken making $k_y = -j\sqrt{k_x^2-k^2}$.

Similar reasoning can be applied to a sheet of magnetic surface current given by
\begin{equation}
	M_x = M_0 e^{-jk_x x}.
\end{equation}
Again, the unique solution which conforms to Maxwell's equations, boundary conditions, symmetry of the problem and the requirement of outward-propagating waves is
\begin{subequations}
	\begin{align}
		E_z(x,y>0) &= -\frac{M_0}{2}e^{-j(k_x x+k_y y)},    &    E_z(x,y<0) &= \frac{M_0}{2}e^{-j(k_x x-k_y y)}, \\
		\mathbf{H}(x,y>0) &= -\frac{M_0}{2\eta k} e^{-j(k_x x+k_y y)} 
		\begin{bmatrix}
			k_y \\
			-k_x \\
			0
		\end{bmatrix}, & 
		\mathbf{H}(x,y<0) &= \frac{M_0}{2\eta k} e^{-j(k_x x-k_y y)} 
		\begin{bmatrix}
			-k_y \\
			-k_x \\
			0
		\end{bmatrix}.
	\end{align}
\end{subequations}

\subsection{Conductive Boundaries} \label{sec:MirageBoundaryConditions}

The metasurfaces considered in this work are defined solely by the boundary conditions they enforce upon the electromagnetic fields surrounding them. Recall from Sec. \ref{sec:HarmonicCurrents} that electric and magnetic surface currents lead to field discontinuities. And so, let the surfaces we consider have electric and magnetic responses. The electric surface response is defined by stating that at a given position, a surface electric current is generated on the surface by a tangential electric field at that position according to
\begin{equation}
	J_z(x,y=0)=\sigma(x)E_z(x,y=0),
\end{equation}
which is defined similarly to the effect of bulk conductivity of a medium in Maxwell's equations, with the main difference that here the $J$ is a surface current density (units of A/m) located on the metasurface and $\sigma$ is a surface conductivity (units of $\Omega^{-1}$). Note that we may drop some or all functional dependence below for brevity. Since an electric surface current produces a discontinuity in tangential magnetic fields without affecting the continuous nature of $E_z$, the term $E_z(x,y=0)$ is perfectly well defined. 

The magnetic response of the surface is defined similarly, and is given by
\begin{equation}
	M_x(x,y=0)=\tau(x)H_x(x,y=0),
\end{equation}
where $M$ is the surface magnetic current on the surface and $\tau$ is a magnetic conductivity of the surface. Similarly, since a magnetic surface current produces a discontinuity in tangential electric fields, the term $H_x(x,y=0)$ is perfectly well defined. Note that the current components shown above are the only non-zero components for the case of 2D, TE fields.

An issue arises when both $\sigma$ and $\tau$ conditions are collocated together as a single metasurface. In this case both $J$ and $M$ exist, and both $E$ and $H$ are discontinuous across the surface. This renders meaningless the terms $E_z(x,y=0)$ and $H_x(x,y=0)$. In order to have a mathematically solvable system of boundary conditions a new rule has to be introduced for how the currents depend upon the fields. We say that in the presence of such a discontinuity the $J$ and $M$ depend upon the average of the fields above and below the surface \cite{kuester2003averaged}. For example, this redefines the electric boundary condition as:
\begin{equation}	
	J_z(x,0)=\sigma(x) \frac{E_z(x,0^+) + E_z(x,0^-)}{2},
\end{equation}
where $E_z(x,0^\pm) = \lim_{\epsilon \to 0} E_x(x,\pm|\epsilon|)$. And likewise for the magnetic response. The validity of the above averaging was also investigated in \cite{kuester2003averaged}. When one considers metasurfaces from a more physical perspective -- being composed of individual scatterers which exhibit electric and magnetic polarizabilities (i.e. meta-atoms) -- it is possible to show that the averaged conductive boundary conditions are a valid approximation, with the error being proportional to the size of the meta-atoms and their spacing. However, in the scope of the theory presented here, there is no way to reason out the averaged boundary conditions. Thus, they must be treated as axioms.

\section{Scattering from Periodic Surfaces via Floquet Theory}

Let us start by noting that constant $\sigma$ and $\tau$ with respect to $x$ do not produce refraction. If one were to try and solve for the scattering from constant conductivities, they would quickly arrive at the conclusion that there will be a transmitted wave and a reflected one, but all with the same magnitudes of $k$-vector components (reflection will follow Snell's law). 

The next natural step in increasing the complexity of the $\sigma\tau$ surface is to follow the theory of diffraction gratings -- it is well known that periodic perturbations in material parameters along some direction cause an incident plane wave to scatter into various other plane waves all having different $k$-vectors. 

And so, consider a periodic $\sigma\tau$ surface of period $T \equiv 2\pi/k_s$. It is well-known that periodic structures impose a quasi-periodicity upon the scattered fields under plane-wave incidence, known as Floquet periodicity \cite{ishimaru2017electromagnetic}. For our problem at hand, Floquet theory dictates that the electric and magnetic surface currents excited on a periodic $\sigma\tau$ surface must be of the form
\begin{align} \label{eq:QuasiPeriodicCurrents}
	J_z(x) &= \sum_{a=-\infty}^\infty J_a e^{-j(k_x +k_s a)x}, & M_x(x) &= \sum_{a=-\infty}^\infty M_a e^{-j(k_x +k_s a)x},
\end{align}
where $J_a$ and $M_a$ are some complex coefficients. According to Sec. \ref{sec:HarmonicCurrents}, each term in the two sums produces TE-polarized plane waves propagating away from the surface in both half-spaces (above and below). The wave-vectors of the produced plane waves are $\bm{k}_a=(k_x+k_s a,k_{y,a},0)$, where $k_{y,a} = \pm \sqrt{k^2 - (k_x+k_s a)^2}$ (see Sec. \ref{sec:HarmonicCurrents} for a discussion regarding the ``$\pm$"). Note that in the future we will also refer to $k_{y,a}$ as $k_a$ (and $k_y$ as $k_0$). 

Let us mention that due to the electromagnetic boundary conditions, the plane waves created by the electric conductivity have zero average tangential magnetic field on the surface, and thus do not induce any magnetic current. The same can be said of the plane waves created by the magnetic conductivity. In a sense, the two conductivities are uncoupled and an independent treatment of each is possible. 

Finally, given the ``Floquet currents" of (\ref{eq:QuasiPeriodicCurrents}) and the discussion of harmonic current sheets of Sec. \ref{sec:HarmonicCurrents}, the total electric field that exists above the metasurface can be written as
\begin{equation} \label{eq:FloquetEFieldsAbove}
	E_z(x,y>0) = E_i e^{-j(k_x x + k_y y)} + E^J_z(x,y>0) + E^M_z(x,y>0),
\end{equation}
where $E^J$ and $E^M$ are the fields created by the electric and magnetic surface currents respectively and have the form
\begin{equation}
	E^{J/M}_z(x,y>0) = \sum_a E^{J/M}_a e^{-j(k_x + k_s a)x -j k_a y}, 
\end{equation}
where $E^{J}_a$ and $E^M_a$ are plane wave amplitudes of the fields created by the $J_a e^{-j(k_x+k_s a)x}$ and $M_a e^{-j(k_x+k_s a)x}$ currents respectively. The magnetic field above the surface can be easily found based on the equations of Sec. \ref{sec:HarmonicCurrents}. 

For the fields below the metasurface a similar equation to (\ref{eq:FloquetEFieldsAbove}) can be written. The incident field is the same and the terms $E^{J/M}_z(x,y<0)$ can be easily obtained from equations of Sec. \ref{sec:HarmonicCurrents}. Again, the magnetic field below the surface can be written easily. This means that the  $E^{J/M}_a$ amplitudes dictate what the total fields above and below the surface have to be. And so, the problem of finding fields which satisfy the $\sigma\tau$ boundary conditions can be written solely in terms of the $E^{J/M}_a$ amplitudes.

\subsection{Tangent Surfaces Subject to Designed Illumination} \label{sec:UniqueScatteringTanSigma}

We know from \cite{selvanayagam2016engineering} that a surface impedance of $1/\sigma \equiv Z = -j\frac{\eta k}{2k_1}\cot(k_s x/2)$ achieves the desired electric response needed for refraction, and so let us consider the surface with a $\sigma(x)$ of
\begin{equation}
	\sigma(x)=j\frac{2k_1}{\eta k} \tan\left(\frac{k_s x}{2}\right) = \frac{2k_1}{\eta k} \left( \frac{e^{j k_s x/2}-e^{-j k_s x/2}}{e^{j k_s x/2}+e^{-j k_s x/2}}  \right).
\end{equation}
The Floquet equations for this surface are obtained via $J_z=\sigma E_z$ by using the above tangent (expressed using the complex exponentials) and the Floquet decomposition of $J_z$ and $E_z$ to obtain (recall also that $J_a=-\frac{2k_a}{\eta k}E^J_a$):
\begin{equation}
	\sum_a -\frac{2k_a}{\eta k}E_a^J e^{-j(k_x +k_s a)x} = \frac{2k_1}{\eta k} \frac{e^{j k_s x/2}-e^{-j k_s x/2}}{e^{j k_s x/2}+e^{-j k_s x/2}}  \left( E_i e^{-jk_x x} +\sum_b E_b^J e^{-j(k_x +k_s b)x} \right).
\end{equation}
Let us now cancel the common $e^{-jk_x  x}$ and multiply both sides of the equation by $e^{jk_s x/2}(e^{jk_s x/2} + e^{-jk_s x/2})\eta k /2 =(e^{jk_s x} + 1)\eta k /2$ to obtain
\begin{equation}
	(1+e^{j k_s x})\sum_a k_a E_a^J e^{-jk_s ax} = k_1 (1-e^{j k_s x})  \left( E_i +\sum_b E_b^J e^{-jk_s bx} \right).
\end{equation}
The $(1\pm e^{j k_s x})$ terms can be absorbed into the summations as
\begin{equation}
	\sum_a (k_a E_a^J e^{-jk_s ax}+k_a E_a^J e^{-jk_s (a-1)x})-\sum_b (k_1 E_b^J e^{-jk_s bx} - k_1 E_b^J e^{-jk_s (b-1)x}) = k_1 (1-e^{j k_s x})  E_i.
\end{equation}
Due to the ``dummy" nature of the $a$ and $b$ indices, we can combine the two summations as
\begin{equation}
	\sum_a \left((k_a E_a^J-k_1 E_a^J) e^{-jk_s ax}+(k_a E_a^J  + k_1 E_a^J) e^{-jk_s (a-1)x}\right) = k_1 (1-e^{j k_s x})  E_i.
\end{equation}
Introducing a new dummy index $c=a-1$ into the $\sum_a (k_a E_a^J  + k_1 E_a^J) e^{-jk_s (a-1)x}$ terms changes their appearance to $\sum_c (k_{c+1} E_{c+1}^J  + k_1 E_{c+1}^J) e^{-jk_s c x}$. This is plugged into the above equation, and again renaming the dummy index $c$ to $a$ gives
\begin{equation}
	\sum_a \left( (k_{a+1}+k_1)E^J_{a+1} + (k_{a}-k_1) E^J_a \right) e^{-j k_s ax} = k_1 (1-e^{jk_s x})E_i.
\end{equation}
A system of equations can be extracted from the above via the orthogonality of complex exponentials, and is written as 
\begin{equation} \label{eq:MatrixFloquetIdealTangentSigma}
	\begin{pmatrix}
		& & & & \vdots & & & & \\
		& k_3+k_1 &  k_2-k_1  &  0  & 0  & 0 & 0 & 0 & \\
		& 0 &  k_2+k_1 &  0  &  0  & 0  & 0 & 0 & \\
		\cdots & 0 & 0 &  2k_1  &  k_0-k_1  &  0  & 0 & 0 & \cdots \\
		& 0 & 0 &  0  & k_0+k_1  &  k_{\text{-}1}-k_1  & 0 & 0 & \\
		& 0 & 0 & 0 &  0  & k_{\text{-}1}+k_1  &  k_{\text{-}2}-k_1  & 0 & \\
		& & & & \vdots & & & & \\
	\end{pmatrix}
	\begin{pmatrix}
		\vdots \\
		E^J_2 \\
		E^J_1 \\
		E^J_0 \\
		E^J_{\text{-}1} \\
		E^J_{\text{-}2} \\
		\vdots \\
	\end{pmatrix} = k_1 E_i
	\begin{pmatrix}
		\vdots \\
		0\\
		0 \\
		1 \\
		-1 \\
		0\\
		\vdots \\
\end{pmatrix}.
\end{equation}
Note that the term $k_1-k_1=0$ (which appears in the first equation above the central one) decouples the system of equations into two independent systems -- one for $E^J_{a>1}$ and the other for $E^J_{a \leq 1}$. In a unique fashion one immediately obtains that $E^J_2=0$. Using this in the next equation above we also obtain $E^J_3=0$. Proceeding to solve the equations via this upward ``step-ladder" argument we obtain $E^J_{a>1} = 0$. 

As for the $E^J_{a\leq 1}$, it first appears that no unique solution exists. One can arbitrarily choose $E^J_1$ and continue the downward step-ladder to find all the other amplitudes. Again we arrive at the fact that the Floquet system alone does not provide us with unique scattering from a surface, and a particular solution must be chosen carefully. After some algebra, the lower mode amplitudes can be written as
\begin{subequations}
\begin{align}
	E^J_0 &= \frac{k_1}{k_0-k_1}\left(E^J_i-2E^J_1\right), \\
	E^J_{-a<0} &= \frac{2k_1(-1)^{a-1}}{(k_{-a}-k_1)(k_0-k_1)}\left( \prod_{n=-a+1}^{-1} \frac{k_n+k_1}{k_n-k_1} \right) \left( (k_0+k_1)E^J_1-k_0E_i \right),
\end{align}
\end{subequations}
where for the case of $E^J_{-1}$, the product term in brackets must be interpreted as 1. In the above, $E^J_1$ is a free parameter which can have any value. And so, again the Floquet system on its own does not provide us with a unique solution to our scattering problem. At this point careful consideration has to be made to narrow the solutions down to a single possibility. 

Let us write the scattered electric field at the surface as
\begin{subequations}
\begin{align}
	E_z^J(x,y=0)-E_i(x,y=0) &= \sum_a E^J_a e^{-j(k_x+k_s a)x} \\
	&= \sum_{-a=-\infty}^{-b \ll -1} E^J_{-a} e^{-j(k_x-k_s a)x}+\sum_{-a=-b+1}^{1}E^J_{-a} e^{-j(k_x-k_s a)x},
\end{align}
\end{subequations}
where the fact that $E^J_{>1}=0$ was used, and $b$ is a very large but finite integer. Let us now ask about the convergence of this expression. The second summation has a finite number of terms, all of which are of finite magnitude, and so the summation itself is a finite in magnitude complex number. However, the same cannot be said of the first summation, which has an infinite number of terms, and so its convergence dictates the convergence of the field. Plugging in the available expression for $E^J_{<0}$ this sum becomes
\begin{equation}
	\sum_{-a=-\infty}^{-b \ll -1} \frac{2k_1(-1)^{a-1}}{(k_{-a}-k_1)(k_0-k_1)}\left( \prod_{n=-a+1}^{-1} \frac{k_n+k_1}{k_n-k_1} \right) \left( (k_0+k_1)E^J_1-k_0E_i \right)e^{-j(k_x-k_s a)x}.
\end{equation}
Manipulating the above expression results in 
\begin{equation} \label{eq:IdealRefFieldConvergence}
	\sum_{-a=-\infty}^{-b \ll -1} E^J_{-a} e^{-j(k_x-k_s a)x} = A(x) \sum_{-a=-\infty}^{-b \ll -1} \frac{(-1)^{a}}{(k_{-a}-k_1)}\left( \prod_{n=-a+1}^{-b} \frac{k_n+k_1}{k_n-k_1} \right) e^{jk_s ax}, \\
\end{equation}
where the common factor $A(x)$ is taken out of each term in the sum, which is
\begin{equation}
		A(x) = -\frac{2k_1}{k_0-k_1}\left( \prod_{n=-b+1}^{-1} \frac{k_n+k_1}{k_n-k_1} \right)\left((k_0+k_1)E^J_1-k_0E_i\right) e^{-jk_x x}.
\end{equation}
Note that this $A(x)$ is an elementary complex function of $x$. For large $a$, $k_{-a} \approx -j k_s a + j k_x$. Using this in the above and taking the limit of infinite $b$ (while keeping in mind that although $b$ goes to infinity, the sum appearing inside the limit has infinite number of terms at every ``step" of the limiting procedure) we can write 
\begin{equation} \label{eq:FieldConvDesInc}
	\lim_{\substack{b \to \infty \\ b < \infty}} \sum_{-a=-\infty}^{-b \ll -1} E^J_{-a} e^{-j(k_x-k_s a)x} = B(x) \cdot \lim_{\substack{b \to \infty \\ b < \infty}} \sum_{-a=-\infty}^{-b\ll -1}\left(\frac{(-e^{jk_sx})^a}{a} \prod_{n=-a+1}^{-b}\frac{j(k_x+k_s n) + k_1}{j(k_x+k_s n)-k_1}\right),
\end{equation}
where $B(x)$ is another elementary function and is given by $B(x)=jA(x)/k_s$. And so, the question whether the fields converge to well-defined values has been transformed into the question whether the sum on the right-hand side of the above converges. First, let us consider the product $\prod_{n=-a+1}^{-b}\frac{j(k_x+k_s n) + k_1}{j(k_x+k_s n)-k_1}$. As $b$ increases, the individual product terms tend to 1. Unfortunately, the same cannot be said of the whole product, since it is composed of (up to) infinitely many terms -- a careful analysis is warranted. For now, let us state that one can show (given a certain interpretation) that $\lim_{\substack{b \to \infty \\ b < \infty}}\left| \prod_{n=-\infty}^{-b} \frac{j(k_x+k_s n) + k_1}{j(k_x+k_s n)-k_1} \right|=1$. With this result, we can crudely argue that in the limit of large $b$ the behavior of the overall sum approaches the behavior of $\lim_{\substack{b \to \infty \\ b < \infty}} \sum_{-a=-\infty}^{-b} \frac{e^{j\phi_a}}{a}$, where $\phi_a$ is some phase. An infinite sum of terms whose magnitudes decrease as $1/a$ has a questionable convergence status -- a sum of the form $\sum_{a \gg 1}^\infty e^{j\phi_a}/a$ may converge or diverge depending on the choice of the phases $\phi_a$. For example, for $\phi_a=0$ the sum becomes $\sum_{a \gg 1}^\infty 1/a$, which is a well-known sum that does not converge. But for $\phi_a=\pi a$, the sum becomes $\sum_{a \gg 1}^\infty (-1)^a/a$, which converges since the terms alternate in sign and approach zero. Let us state that indeed there are positions along the surface at which the sum of (\ref{eq:FieldConvDesInc}) does not converge (the detailed proof of this statement is provided at the end of this document separately). And so, we find that if any plane wave amplitude $E^J_{<0}$ is non-zero, there will be points on the surface at which the fields do not converge (i.e. have no value) -- an unphysical and unwanted situation. Based on physical arguments these solutions must be thrown out, and the only way to achieve this is to set $A(x)$ to 0. This, together with the expression for $E^J_0$ gives us a unique physical solution to the scattering problem: 
\begin{align}
	E^J_0 &= -\frac{k_1}{k_0+k_1} E_i,  &  E^J_1 &= \frac{k_0}{k_0+k_1}E_i, & E^J_{a\neq 0,1}&=0.
\end{align}

In summary, the tangent $\sigma$ considered here is capable of achieving the electric response needed for refraction. Although this is not a new result in itself, this section studies the subtle question of whether the solution to the Floquet scattering problem is unique. Indeed, uniqueness is established when non-physical solutions to the Floquet system are thrown out. Note however, that this was shown to be true only for the case of designed incidence. The same cannot be said of general incidence, as we will see below. 

Turning our attention to a $\tau$-only surface, the surface achieving the required magnetic response is given by 
\begin{equation}
	\tau(x) = j\frac{2\eta k}{k_1}\tan\left(\frac{k_sx}{2}\right).
\end{equation}
Again, mirroring the analysis above, a Floquet system can be written and similar arguments can be made regarding finding a physically unique solution to the scattering. The scattering solution is given by
\begin{align}
	E^M_0 &= -\frac{k_0}{k_0+k_1} E_i, & E^M_1 &= \frac{k_0}{k_0+k_1}E_i, & E^M_{a\neq 0,1}&=0.
\end{align}

Placing the $\sigma$ and $\tau$ together achieves refraction for designed incidence -- the electric and magnetic responses and the incident fields add to produce only the refracted wave above the surface and the incident and reflected waves below. 

\subsection{Tangent Surfaces Subject to General Illumination} \label{sec:IlldefinedScattering}

In the previous section we have used the well-known tangent (commonly referred to as cotangent) expressions which achieve refraction, and solved the scattering problem in a much more rigorous manner than existing solutions. However, the scattering problem was solved in an incomplete fashion -- scattering was deduced only in the case when the illuminating plane wave is incident at the particular designed angle (given by $\atan(k_x/k_0)$). Let us now ask what the scattering will be when a more general plane wave illuminates these tangent surfaces. 

Notationally, general incidence will be encompassed by priming the wave vectors of the incident and scattered plane waves. The wave vector of the incident plane wave is written as ${\bm k}_i' = k_x' \hat{\bm x}+k_0'\hat{\bm y}$, and that of the various Floquet modes as ${\bm k}_a' = (k_x'+k_s a) \hat{\bm x}+k_a'\hat{\bm y}$. For now, let us study the scattering at the single frequency for which the surface was ``designed", making $k'=k$.  Keep in mind that the surface parameters are still defined to achieve refraction between the unprimed ${\bm k}_i \to {\bm k}_1$. Thus, the expressions defining $\sigma$ and $\tau$ still contain the unprimed $k_1$. With these comments in mind, the Floquet system for the $\sigma$-only surface can be written as
\begin{equation}
	\sum_a \left((k_{a+1}' +k_1)E^J_{a+1} + (k_a'-k_1) E^J_a \right) e^{-jk_s a x} = k_1 \left(1-e^{jk_s x}\right)E_i,
\end{equation}
which in matrix form becomes
\begin{equation} 
	\begin{pmatrix}
		& & & & \vdots & & & & \\
		& k_3'+k_1 &  k_2'-k_1  &  0  & 0  & 0 & 0 & 0 & \\
		& 0 &  k_2'+k_1 &  k_1'-k_1  &  0  & 0  & 0 & 0 & \\
		\cdots & 0 & 0 &  k_1'+k_1  &  k_0'-k_1  &  0  & 0 & 0 & \cdots \\
		& 0 & 0 &  0  & k_0'+k_1  &  k_{\text{-}1}'-k_1  & 0 & 0 & \\
		& 0 & 0 & 0 &  0  & k_{\text{-}1}'+k_1  &  k_{\text{-}2}'-k_1  & 0 & \\
		& & & & \vdots & & & & \\
	\end{pmatrix}
	\begin{pmatrix}
		\vdots \\
		E^J_2 \\
		E^J_1 \\
		E^J_0 \\
		E^J_{\text{-}1} \\
		E^J_{\text{-}2} \\
		\vdots \\
	\end{pmatrix} = k_1 E_i
	\begin{pmatrix}
		\vdots \\
		0\\
		0 \\
		1 \\
		-1 \\
		0\\
		\vdots \\
	\end{pmatrix}.
\end{equation}
Note that because $k_1' \neq k_1$, decoupling of the system of equations does not occur. Let us assume that the scattering amplitudes of the central modes (say the $E^J_{0,1}$) are not infinite, and at least one is non-zero. In this case, the solutions to $E^J_{>1}$ and $E^J_{<0}$ can be written in terms of $E^J_{0,1}$ as was done previously. Performing the same analysis, one obtains
\begin{subequations} \label{eq:IdealRefFieldConvergence2}
\begin{align}
	E^J_{a\gg 1} &= A(x) \frac{(-1)^a}{k_a'+k_1}\prod_{n=b}^{a-1} \frac{k_n'-k_1}{k_n'+k_1} \hphantom{a}\to\hphantom{a} B(x)\frac{e^{j\phi_a}}{a},\\
	E^J_{-a\ll -1} &= A(x) \frac{(-1)^a}{k_{-a}'-k_1}\prod_{n=-a+1}^{-b} \frac{k_n'+k_1}{k_n'-k_1} \hphantom{a}\to\hphantom{a} B(x)\frac{e^{j\psi_a}}{a},
\end{align}
\end{subequations}
where again $a \gg b \gg 1$, $A(x)$ and $B(x)$ are some elementary complex functions ($A$ and $B$ are used as placeholders, are not equal to the previous definitions, nor among themselves in (\ref{eq:IdealRefFieldConvergence2}a) and (\ref{eq:IdealRefFieldConvergence2}b)), and $\phi_a$ and $\psi_a$ are some phases. The above amplitudes have the same form as that of the non-converging fields considered in the previous section. And so, here too, the fields do not converge along the surface unless all of the $E^J_{|a|\gg 1}$ modes are zero (again, this statement will be proven rigorously later on). Let us see what this ``physicality" condition leads to for the values of the central Floquet modes. Via step-ladder arguments from the above and below (first start with some $E^J_{a\gg 1}=0$ and proceed solving downwards, then start with some $E^J_{a\ll-1}=0$ and proceed solving upwards, finally around the central equations the two ``step-ladders" meet), the system of equations tells us that $E^J_{a\neq 0}=0$, while $E^J_0$ must conform to the two equations which read
\begin{align}
	E^J_0 &= \frac{k_1}{k_0'-k_1}E_i, & E^J_0 &= -\frac{k_1}{k_0'+k_1}E_i,
\end{align}
which has any meaning only if $k_0'=0$, implying that ${\bm k}_i' = \pm k\hat{\bm x}$. And so, for the case of general incidence (wave vector ${\bm k}_i' \neq {\bm k}_i, \pm k\hat{\bm x}$) there does not seem to be a physically valid solution to the scattering.

Similar behavior can be observed by performing the same analysis on the $\tau$-only surface. These findings depict that the imaginary tangent variations for $\sigma$ and $\tau$ constitute scattering problems which generally do not possess a physically valid solution (under non-designed incidence). 

\subsection{Lossy Tangents Subject to Designed Incidence}

Let us now fix the non-convergence of the fields which appeared in non-designed incidence upon lossless tangent $\sigma$ and $\tau$ surfaces. One way to force uniqueness upon the Floquet system of equations (in the sense that all possible solutions of the infinite system of equations can be thrown out except one based on physical arguments), as will be shown below, is to introduce loss into the originally lossless tangent.

\subsubsection{$\sigma$-only Case}

Let us first consider only the electric surface response, and think about how to alter the lossless tangent to lead to a solvable scattering problem. 
At first glance, for the $\sigma(x)=j\frac{2k_1}{\eta k} \tan\left(\frac{k_s x}{2}\right) = j\frac{2k_1}{\eta k} \frac{\sin(k_s x/2)}{\cos(k_s x/2)}$ mathematical problems may arise when the cosine in the denominator becomes zero. Let us first attempt to get rid of divisions by zero at any location on the surface by some small lossy perturbation. Thus we should make sure that given any non-zero perturbation, the denominator never becomes zero. Adding some real number in the denominator (i.e. $\cos(k_s x/2) \to \cos(k_s x/2)+\Delta$) does not fix the issue, since this doesn't necessarily remove the singularities (it simply offsets them if $|\Delta|\le 1$). This is unfortunate, because had this perturbation worked, the surface would remain lossless and passive. If instead, one adds a purely imaginary number (i.e. $\cos(k_s x/2) \to \cos(k_s x/2)+j\Delta$) the singularities do disappear, but this form of $\sigma(x)$ has problems. Firstly, it doesn't lead to a nice-looking Floquet system of equations. But more importantly, such perturbed $\sigma$ would have a negative real part at some locations which corresponds to an active surface. And so, we choose the perturbed $\sigma$ to have the form
\begin{equation}
	\sigma_\Delta(x) = j\frac{2k_1}{\eta k} \frac{\sin(k_s x/2)}{\cos(k_s x/2)+j\Delta\sin(k_s x/2)},
\end{equation}
where $\Delta$ is a small (but finite) positive number. Note that the denominator indeed never becomes zero. Under this perturbation the resulting Floquet system is simple and the surface is always passive ($\Re{\sigma(x)}\ge 0$, as can be easily verified). Using the complex exponential expansion of the cosines and sines, and using the same approach that appears above, the Floquet equations for this surface can be written as
\begin{equation} \label{eq:FloquetSigmaDeltaLoss}
	\sum_a \left( (k_{a+1}(1+\Delta)+k_1)E^J_{a+1} +(k_a(1-\Delta)-k_1)E^J_a\right)e^{-jk_s a x} = k_1 (1-e^{j k_s x}) E_i,
\end{equation}
and in matrix form this becomes
\begin{equation}
	\begin{pmatrix}
		& & \vdots & & \\
		   &  0  & 0  & 0 &  \\
		  &  k_1(1-\Delta)-k_1  &  0  & 0  &  \\
		\cdots &   k_1(1+\Delta)+k_1  &  k_0(1-\Delta)-k_1  &  0  &   \cdots \\
		  &  0  & k_0(1+\Delta)+k_1  &  k_{\text{-}1}(1-\Delta)-k_1  &   \\
		  & 0 &  0  & k_{\text{-}1}(1+\Delta)+k_1  &    \\
		& & \vdots & & \\
	\end{pmatrix}
	\begin{pmatrix}
		\vdots \\
		E^J_2 \\
		E^J_1 \\
		E^J_0 \\
		E^J_{\text{-}1} \\
		E^J_{\text{-}2} \\
		\vdots \\
	\end{pmatrix} = k_1 E_i
	\begin{pmatrix}
		\vdots \\
		0\\
		0 \\
		1 \\
		-1 \\
		0\\
		\vdots \\
	\end{pmatrix}.
\end{equation}

Let us now proceed in the same manner as previously, and obtain the expressions for $E^J_{a\gg 0}$ and $E^J_{a\ll0}$. Without repeating the same arguments, we write
\begin{subequations}
\begin{align} \label{eq:AsymptoticEJaLossySigma}
	E^J_{a\gg 1} &= A(x) \frac{(-1)^a}{(1+\Delta)k_a+k_1}\prod_{n=b}^{a-1} \frac{(1-\Delta)k_n-k_1}{(1+\Delta)k_n+k_1} \hphantom{a}\to\hphantom{a} B(x)\frac{e^{j\phi_a}}{a} \prod_{n=b}^{a-1} \frac{1-\Delta}{1+\Delta},\\
	E^J_{-a\ll -1} &= A(x) \frac{(-1)^a}{(1-\Delta)k_{-a}-k_1}\prod_{n=-a+1}^{-b} \frac{(1+\Delta)k_{n}+k_1}{(1-\Delta)k_{n}-k_1} \hphantom{a}\to\hphantom{a} B(x)\frac{e^{j\psi_a}}{a}\prod_{n=-a+1}^{-b} \frac{1+\Delta}{1-\Delta},
\end{align}
\end{subequations}
where $A(x)$ and $B(x)$ are again just placeholders for bounded elementary complex functions of $x$ (which are not necessarily equal to the $A(x)$ and $B(x)$ appearing previously or even among each other in the above two equations) and $\phi_a$ and $\psi_a$ are again some generic phases (also not equal to the previously appearing phases). From the above, one can deduce that the convergence of the fields depends on the independent convergence of two sums of the form
\begin{align} \label{eq:FieldConvergenceSumsLossySigma}
	\sum_{a\gg 1}^\infty \frac{e^{j\phi_a}}{a}\left( \frac{1-\Delta}{1+\Delta} \right)^a, && \sum_{-a=-\infty}^{-a \ll -1} \frac{e^{j\psi_a}}{a}\left( \frac{1+\Delta}{1-\Delta} \right)^a.
\end{align}
Because $\Delta$ is positive and finite, $\frac{1-\Delta}{1+\Delta}<1$ and $\frac{1+\Delta}{1-\Delta}>1$. Noting that 
\begin{align}
	\sum_{a\gg 1}^\infty \left| \frac{e^{j\phi_a}}{a}\left( \frac{1-\Delta}{1+\Delta} \right)^a \right| &< \sum_{a\gg 1}^\infty \left( \frac{1-\Delta}{1+\Delta} \right)^a, & \sum_{-a=-\infty}^{-a \ll -1} \left| \frac{e^{j\psi_a}}{a}\left( \frac{1+\Delta}{1-\Delta} \right)^a\right| > \sum_{-a=-\infty}^{-a \ll -1} \frac{1}{a},
\end{align}
we immediately see that the left sum converges, and the one on the right diverges. Since the right sum was obtained by considering the $E^J_{\ll -1}$ amplitudes, these amplitudes must be zero to avoid non-convergence of the field. On the other hand, the left sum is due to the $E^J_{\gg 1}$ amplitudes, and the converging nature of the sum shows that these amplitudes do not have to be zero.

Using the above deductions, we can analytically solve for the scattering of the lossy tangent that is subject to designed incidence. Since $E^J_{\ll -1}=0$, we can use the upward step-ladder argument to obtain
\begin{subequations}
\begin{align}
	E^J_{<0} &= 0, \\
	E^J_{0} &= -\frac{k_1}{(1+\Delta)k_0+k_1}E_i, \\
	E^J_{1} &= \frac{1}{1+\Delta/2}\cdot\frac{k_0}{(1+\Delta)k_0+k_1}E_i, \\	
	E^J_{a>1} &= \frac{(-1)^{a+1}}{(1+\Delta)k_a+k_1}\left( \prod_{n=1}^{a-1} \frac{(1-\Delta)k_n-k_1}{(1+\Delta)k_n+k_1} \right) \frac{2k_0 k_1}{(1+\Delta)k_0+k_1}E_i.
\end{align}
\end{subequations}

\subsubsection{$\tau$-only Case}

Let us now turn our attention to a lossy tangent-like $\tau$-only surface. Using analogous perturbation, we write
\begin{equation}
	\tau_\Delta(x) = j\frac{2\eta k}{k_1} \frac{\sin(k_s x/2)}{\cos(k_s x/2)+j\Delta\sin(k_s x/2)}.
\end{equation}
Using this magnetic response, the Floquet system of equations becomes
\begin{equation}
	\sum_a \left(  (k_1(1+\Delta)+k_{a+1})E^M_{a+1} + (k_1(1-\Delta)-k_a)E^M_a  \right) e^{-j k_s a x} = (1-e^{jk_s x})k_0 E_i.
\end{equation}
Note the key difference between this expression and the electric analog of (\ref{eq:FloquetSigmaDeltaLoss}) is that here the $(1\pm\Delta)$ terms appear together with the $k_1$, as opposed to with $k_{a+1}$ and $k_{a}$. For the lossy $\sigma$-only surface we were able to establish a unique, well-behaving solution because the factors $(1\pm\Delta)$ appeared with $k_{a+1}$ and $k_{a}$. Following the same analysis with the above Floquet system, one will quickly find that it is unsolvable. How can that be, one may ask. After all, again we made sure that the conductivity is well-behaving and never approaches infinity. We will provide an explanation for this below, but for now, let us ask what kind of magnetic loss is required to have a solvable scattering problem. 

And so, we would like a $\tau$ which would place $(1\pm\Delta)$-like terms with $k_{a+1}$ and $k_{a}$. A simple choice of $\tau$ which achieves this is
\begin{equation}
	\tau_\Lambda(x)=j\frac{2\eta k}{k_1}\tan\left(\frac{k_s x}{2}\right) + \frac{2\eta k}{k_1}\Lambda,
\end{equation}
where $\Lambda$ is a small (but finite) positive number (note that we assume $k_1 > 0$). A different symbol is used for this perturbation to signify that this is a different loss mechanism compared to the $\Delta$ term used for $\sigma$ above. With this $\tau_\Lambda$, the Floquet equations become
\begin{equation}
	\sum_a \left(  \left(k_1+(1+\Lambda)k_{a+1}\right)E^M_{a+1}+\left(k_1-(1-\Lambda)k_a\right)E^M_a  \right) e^{-jk_s a x} = \left(  (1-\Lambda)-(1+\Lambda)e^{jk_s x}  \right) k_0 E_i.
\end{equation}
Now that the $(1\pm\Lambda)$ terms appear in the desired locations, which guarantees the existence of a unique well-behaving solution. Following this same analysis as for the $\sigma$-only surface above, the solution to the scattering can be obtained. Without repeating the same arguments, the solution is
\begin{subequations}
\begin{align}
	E^M_{<0} &= 0,\\
	E^M_{0} &= -\frac{k_0}{k_0+\frac{k_1}{1+\Lambda}}E_i, \\
	E^M_{1} &= \frac{1}{1+\Lambda/2}\cdot\frac{k_0}{(1+\Lambda)k_0+k_1}E_i, \\
	E^M_{a>1} &= \frac{(-1)^{a+1}}{k_1+(1+\Lambda)k_a}\left( \prod_{n=1}^{a-1} \frac{k_1-(1-\Lambda)k_n}{k_1+(1+\Lambda)k_n} \right)\frac{2k_0 k_1}{k_1+(1+\Lambda)k_0}E_i.
\end{align}
\end{subequations}

\subsection{$\Delta$- and $\Lambda$-type Loss Mechanisms}

Looking at the previous section, some question arise. We first perturbed $\sigma$ to get rid of the infinities and established that then the problem becomes well-defined. However, doing the same to $\tau$, did not achieve this. Furthermore, the $\tau_\Lambda$ exhibits infinities, which doesn't seem to effect the well-behaving nature of the scattering. Finally, from a visual perspective, the similarly-defined $\sigma$ and $\tau$ appear somewhat symmetrically in Maxwell's equations. Why do the required loss perturbations appear differently, ``non-symmetrically"? 

The answer to this lies in the fact that up to now we were only discussing TE-polarized fields. Let us quickly investigate what happens with the $\sigma$-only surface when subject to a TM-polarized incident wave. Now, $H_z \neq 0$ wile $H_x=H_y=0$. Instead of referring to plane waves by their electric field amplitudes ($E_a$), the magnetic field amplitudes are more natural. And so, let us now write the Floquet equations (in terms of $H_a$) for TM incidence upon the $\sigma_\Delta$ surface. In order to do this, one must note that for TM fields $J_a=2H_a$ and $E_x=-H_z \eta k_y/k$. This gives
\begin{equation}
	\sum_a \left(  \left(k^2(1+\Delta) + k_1 k_{a+1} \right) H_{a+1} + \left(k^2(1-\Delta) - k_1 k_{a} \right) H_{a}  \right)e^{-jk_s a x} = (1-e^{jk_s x}) k_0 k_1 H_i.
\end{equation}
Note that the $(1\pm\Delta)$ terms are separate from the $k_{a,a+1}$. And so, by inspection, this scattering problem is unsolvable. This is somewhat odd -- the $\sigma_\Delta$ is always well-behaving and non-infinite. Thus, it appears that simply a well-behaving surface conductivity (electric or magnetic) is not sufficient to guarantee well-behaving scattering for general plane-wave incidence (of any polarization).

One may check that adding a $\Lambda$-type loss mechanism to $\sigma$ leads to a solvable TM incidence case. Thus, it appears that both $\sigma$ and $\tau$ require both types of loss:
\begin{subequations}
\begin{align}
	\sigma_{\Delta\Lambda}(x) &= j\frac{2k_1}{\eta k}\frac{\sin(k_s x/2)}{\cos(k_s x/2)+j\Delta_\sigma\sin(k_s x/2)}+\frac{2}{\eta k}\Lambda_\sigma, \\
	\tau_{\Delta\Lambda}(x) &= j\frac{2\eta k}{k_1}\frac{\sin(k_s x/2)}{\cos(k_s x/2)+j\Delta_\tau\sin(k_s x/2)}+\frac{2\eta}{k_1}\Lambda_\tau,
\end{align}
\end{subequations}
where the subscripts under the $\Delta$ and $\Lambda$ symbolize that the electric and magnetic losses are independent, and the extra factors of $2/\eta k$ and $2\eta/k_1$ were introduced in front of $\Lambda$'s to simplify future Floquet systems. This form of lossy tangents will be used exclusively from now on. This form allows for general plane wave incidence and is symmetric (indeed, as Maxwell's equations dictate).

\subsection{Lossy Tangents Subject to General TE Illumination}

Let us now investigate the scattering of lossy tangent-like surfaces under general plane wave illumination (incident wave vector ${\bm k}'_i$ being completely independent of the original ${\bm k}_i$). First, considering a $\sigma$-only surface, the Floquet equations read
\begin{multline}
	\sum_a \left[  \left( (1+\Delta)\frac{k}{k'}k'_{a+1}+k_1+(1+\Delta)\Lambda \right)E^J_{a+1}+\left( (1-\Delta)\frac{k}{k'}k'_{a}-k_1+(1-\Delta)\Lambda \right)E^J_a   \right]e^{-jk_s ax} \\ = -\left(  (1-\Delta)\Lambda - k_1 + \left( (1+\Delta)\Lambda+k_1 \right) e^{jk_s x}  \right) E_i.
\end{multline}
Using identical arguments as above, the final solution to this system can be found to be
\begin{subequations}
\begin{align}
	E^J_{<0} &= 0, \\
	E^J_0 &= -\frac{k_1+(1+\Delta)\Lambda}{k_1+(1+\Delta)\left(\frac{k}{k'}k'_0+\Lambda\right)}E_i, \\
	E^J_1 &= \frac{2k_1 k'_0 \frac{k}{k'}}{\left( k_1+(1+\Delta)\left(\frac{k}{k'}k'_1 + \Lambda \right) \right) \left( k_1+(1+\Delta)\left(\frac{k}{k'}k'_0 + \Lambda \right) \right)} E_i,\\
	E^J_{a>1} &= \frac{(-1)^{a+1}}{k_1+(1+\Delta)\left(\frac{k}{k'}k'_a+\Lambda\right)} \left( \prod_{n=1}^{a-1} \frac{-k_1+(1-\Delta)\left(\frac{k}{k'}k'_n+\Lambda\right)}{k_1+(1+\Delta)\left(\frac{k}{k'}k'_n+\Lambda\right)} \right)  \frac{2 k_1 k'_0 \frac{k}{k'}}{k_1+(1+\Delta)\left(\frac{k}{k'}k'_0+\Lambda\right)}E_i.
\end{align}
\end{subequations}
Note that in these equations no constraints are placed on the primed wave vector components. Thus, these equations provide the general scattering solution (for the case of TE incidence) from a lossy tangent-like $\sigma$-only surface (which itself provides ``refraction" at a given frequency and incident angle as given by ${\bm k}_i$ and ${\bm k}_1$). 

For a lossy tangent-like $\tau$-only surface, same arguments lead to the Floquet system of
\begin{multline}
	\sum_a \left[  \left( k_1(1+\Delta)+(k+\Lambda(1+\Delta))\frac{k'_{a+1}}{k'} \right)E^M_{a+1}+\left( k_1(1-\Delta)-(k-\Lambda(1-\Delta))\frac{k'_{a}}{k'} \right)E^M_a   \right]e^{-jk_s ax} \\ = \left(  k-\Lambda(1-\Delta) - \left( k+\Lambda(1+\Delta) \right) e^{jk_s x}  \right)\frac{k'_0}{k'} E_i,
\end{multline}
with the solution being
\begin{subequations}
\begin{align}
	E^M_{<0} &= 0, \\
	E^M_0 &= -\frac{k+\Lambda(1+\Delta)}{k_1(1+\Delta)+(k+\Lambda(1+\Delta))\frac{k'_0}{k'}}\frac{k'_0}{k'}E_i, \\
	E^M_1 &= \frac{2 k_1 k'_0 \frac{k}{k'}}{\left(k_1(1+\Delta)+(k+\Lambda(1+\Delta))\frac{k'_1}{k'}\right) \left(k_1(1+\Delta)+(k+\Lambda(1+\Delta))\frac{k'_0}{k'}\right)}E_i, \\
	E^M_{a>1} &= \frac{(-1)^{a+1}}{k_1(1+\Delta)+(k+\Lambda(1+\Delta))\frac{k'_a}{k'}} \left(\prod_{n=1}^{a-1} \frac{k_1(1-\Delta)-(k-\Lambda(1-\Delta))\frac{k'_n}{k'}}{k_1(1+\Delta)+(k+\Lambda(1+\Delta))\frac{k'_n}{k'}}\right) \frac{2k_1 k'_0 \frac{k}{k'} }{k_1(1+\Delta)+(k+\Lambda(1+\Delta))\frac{k'_0}{k'}}E_i.
\end{align}
\end{subequations}
The equations above provide the analytical solution to the scattered fields from a physical tangent-like $\tau$-only surface for any TE incident plane wave.

In conclusion, we would like to mention that in the above equations for $E^J_a$ and $E^M_a$ one can set both $\Delta$ and $\Lambda$ to zero without ruining the well-defined nature of the expressions. And so, we can say that in the limit of zero loss ($\Delta, \Lambda \to 0$) the scattering of the tangent $\sigma$-only and $\tau$-only surfaces reduces to
\begin{subequations} \label{eq:GenScatLimZeroLoss}
\begin{align}
	E^J_{<0} &= 0, & E^M_{<0} &=0, \\
	E^J_0 &= -\frac{k_1}{k_1 +\frac{k}{k'} k'_0}E_i, & E^M_0 &= -\frac{\frac{k}{k'} k'_0}{k_1+\frac{k}{k'} k'_0}E_i, \\
	E^J_1 &= \frac{2k_1 k'_0 \frac{k}{k'}}{\left(k_1 +\frac{k}{k'} k'_1\right)\left(k_1+\frac{k}{k'}k'_0\right)}E_i, & E^M_1 &= \frac{2k_1  k'_0 \frac{k}{k'}}{\left(k_1+\frac{k}{k'} k'_1\right)\left(k_1 +\frac{k}{k'} k'_0\right)}E_i, \\
	E^J_{a>1} &= \frac{(-1)^{a+1}2k_1 k'_0 \frac{k}{k'} E_i}{\left(k_1+\frac{k}{k'} k_a'\right)\left(k_1+\frac{k}{k'} k'_0\right)} \prod_{n=1}^{a-1}\frac{\frac{k}{k'} k'_n-k_1}{\frac{k}{k'}k'_n + k_1}, & E^M_{a>1} &= \frac{(-1)^{a+1}2k_1 k'_0 \frac{k}{k'} E_i}{\left(k_1+\frac{k}{k'}k_a'\right)\left(k_1+\frac{k}{k'} k'_0\right)} \prod_{n=1}^{a-1}\frac{k_1-\frac{k}{k'} k'_n}{k_1 + \frac{k}{k'} k'_n}.
\end{align} 
\end{subequations}
Of course, if one uses the above expressions, they should keep in mind that they are valid only in the limiting sense -- after all, we proved in Sec. \ref{sec:IlldefinedScattering} that the general scattering of a truly lossless tangent-like surface does not converge. Furthermore, it must be stressed that the above equations are only useful for finding approximate values of Floquet mode amplitudes of low mode number, since the overall sum of plane waves given by the above equations diverges at various points on the surface, again as proved in Sec. \ref{sec:IlldefinedScattering}. 

Finally, we remind the reader that although we have explicitly discussed scattering of tangent-like $\sigma$-only and $\tau$-only surfaces, the scattering of a combined $\sigma\tau$ surface is given by the sum of the individual responses.
We also highlight that the obtained scattering amplitudes (both the lossy and limit of zero loss cases) extend current investigations of such surfaces beyond the scattering under non-designed angle of incidence -- the equations inherently allow evaluation at any frequency, thus completely specifying the response of the boundary under 2D TE incidence.

\subsection{Reflection \& Transmission Coefficients}

So far, we have obtained the general 2D TE scattering solution of the ``refracting" $\sigma\tau$ surface by providing closed-form expression for the plane-wave amplitudes $E_a^J$ and $E_a^M$. However, some further insight may be gained by writing the solution in a different form. Let us re-write the fields above and below the boundary as
\begin{subequations}
\begin{align}
	E_z(x,y>0) &= E_i \sum_a \mathrm{T}_a e^{-j(k'_x+k_s a)x-j k'_a y},\\
	E_z(x,y<0) &= E_i \left(e^{-jk'_x x -j k'_0 y}+\sum_a \Gamma_a e^{-j(k'_x+k_s a)x+j k'_a y}\right),
\end{align}
\end{subequations}
where $\mathrm{T}_a$ and $\Gamma_a$ are the transmission and reflection coefficients into the $a^\mathrm{th}$ Floquet mode. It is easy to deduce that, in terms of the known $E^{J/M}_a$ amplitudes, the new coefficients are given by
\begin{subequations}
	\begin{align}
		\mathrm{T}_a &= \delta_{a,0} + \frac{E^J_a+E^M_a}{E_i},\\
		\Gamma_a &= \frac{E^J_a-E^M_a}{E_i},
	\end{align}
\end{subequations}
where $\delta_{a,0}$ is the Dirac delta function.

Let us now provide concrete expressions for $\mathrm{T}_a$ and $\Gamma_a$ in the limit of zero loss ($\Delta,\Lambda \to 0$) and study their behavior. Using (\ref{eq:GenScatLimZeroLoss}) in the above definitions one quickly obtains
\begin{subequations}
	\begin{align}
		\mathrm{T}_{a<0} &= 0, & \Gamma_{a<0} &=0, \\
		\mathrm{T}_0 &= 0, & \Gamma_0 &= -\frac{k_1-\frac{k}{k'} k'_0}{k_1+\frac{k}{k'} k'_0}, \\
		\mathrm{T}_1 &= \frac{4k_1 k'_0 \frac{k}{k'}}{\left(k_1 +\frac{k}{k'} k'_1\right)\left(k_1+\frac{k}{k'}k'_0\right)}, & \Gamma_1 &= 0, \\
		\mathrm{T}_{a>1} &= \frac{(-1)^{a+1}+1}{k_1+\frac{k}{k'} k_a'}\cdot\frac{2k_1 k'_0 \frac{k}{k'}}{k_1+\frac{k}{k'} k'_0} \prod_{n=1}^{a-1}\frac{\frac{k}{k'} k'_n-k_1}{\frac{k}{k'}k'_n + k_1}, & \Gamma_{a>1} &= \frac{(-1)^{a+1}-1}{k_1+\frac{k}{k'} k_a'}\cdot\frac{2k_1 k'_0 \frac{k}{k'}}{k_1+\frac{k}{k'} k'_0} \prod_{n=1}^{a-1}\frac{\frac{k}{k'} k'_n-k_1}{\frac{k}{k'}k'_n + k_1}.
	\end{align} 
\end{subequations}
These are the scattering equations provided in the main text of the letter.

Let us explicitly mention what these equations collapse to under designed incidence ($k'=k$ and $k_x'=k_x$). In this case, it is easy to see that the solution becomes
\begin{subequations}
	\begin{align}
		\mathrm{T}_{a<0} &= 0, & \Gamma_{a<0} &=0, \\
		\mathrm{T}_0 &= 0, & \Gamma_0 &= -\frac{k_1- k_0}{k_1+ k_0}, \\
		\mathrm{T}_1 &= \frac{2 k_0 }{k_1+k_0}, & \Gamma_1 &= 0, \\
		\mathrm{T}_{a>1} &= 0, & \Gamma_{a>1} &= 0,
	\end{align} 
\end{subequations}
showing refraction (an incident wave excites a single refracted and a single reflected waves). This is indeed in agreement with past work of \cite{selvanayagam2016engineering}.

\subsection{Case Studies -- Continued}

Let us now build upon the numerical case studies of the main letter. Again, we set $E_i=1$ and keep $k$ variable.

\begin{figure}[tbh!]
	\centering
	\includegraphics[width=0.6\textwidth]{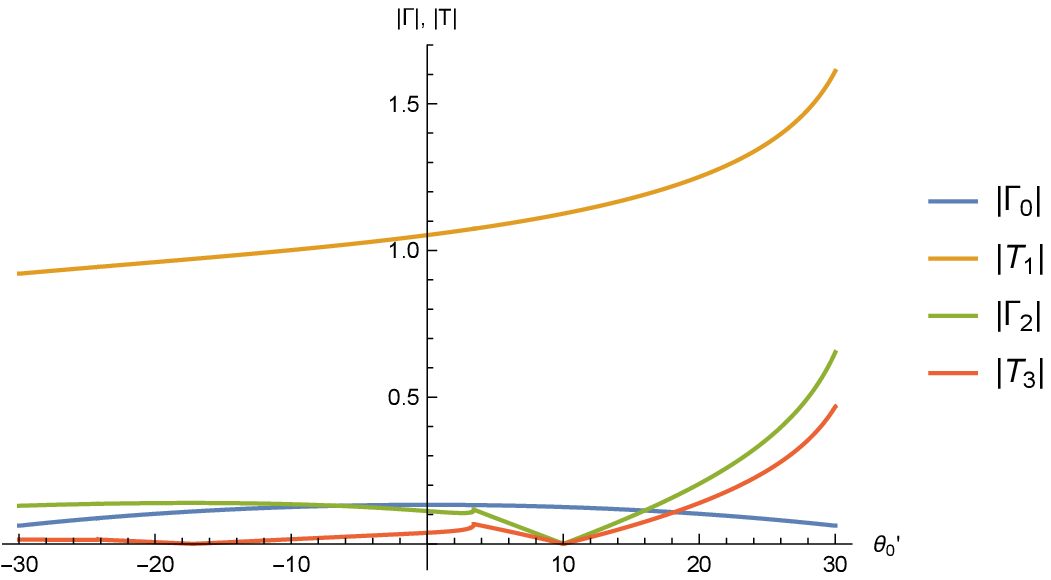}
	\caption{A few $|\mathrm{T}_a|$ and $|\Gamma_a|$ versus incident angle $\theta'_0$ (at the design frequency) for a lossless surface with $k_s=0.47k$ and $k_x=0.174k$.}
	\label{fig:SelvComp}
\end{figure}

We now study a surface with a peculiar choice of parameters -- 
$k_s = 0.47 k$ and $k_x = 0.174 k$ (for such a surface $\theta_0=10^\circ$ and $\theta_1=40^\circ$). In fact, the scattering of this surface was numerically investigated by Selvanayagam in \cite{selvanayagam2016engineering}. Figure \ref{fig:SelvComp} shows the magnitude of a few $\mathrm{T}$ and $\Gamma$ coefficients versus incidence angle (at the design frequency). The curves in this figure replicate the data in Fig. 6.7(a)-(b) of \cite{selvanayagam2016engineering} -- in fact, Selvanayagam's numerical studies agree with our analytical solution. At this point, we explicitly mention that our solution does not agree with Epstein's solution \cite{epstein2014floquet} -- we found that negative mode amplitudes ($\mathrm{T}_{a<0}$ and $\Gamma_{a<0}$) have to be zero, while for Epstein they are not (see for example Fig. 4 of \cite{epstein2014floquet}).

\begin{figure}[tbh!]
	\centering
	\includegraphics[width=0.6\textwidth]{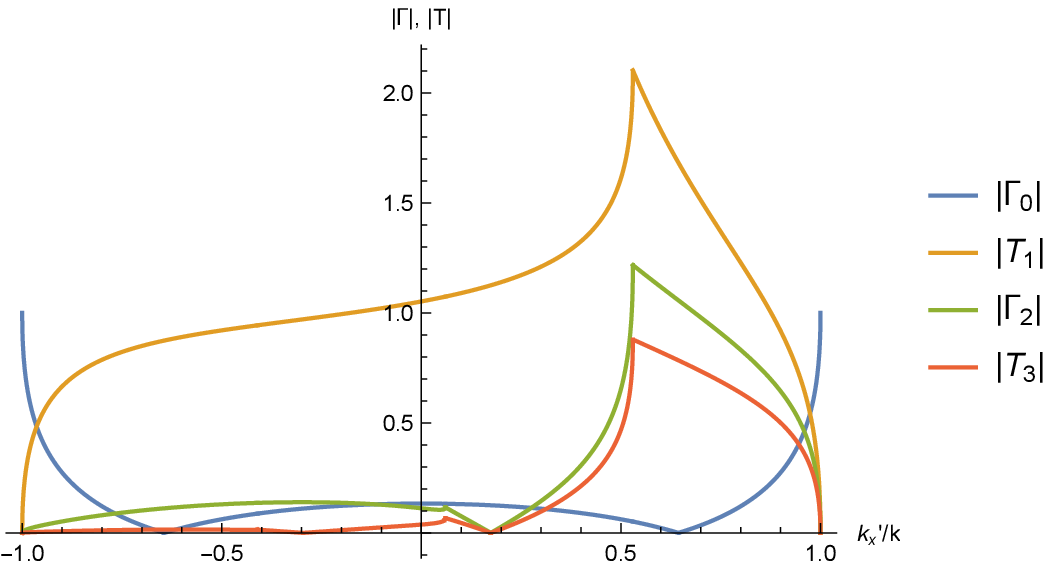}
	\caption{A few $|\mathrm{T}_a|$ and $|\Gamma_a|$ versus incident angle $k'_x/k$ (at the design frequency) for a lossless surface with $k_s=0.47k$ and $k_x=0.174k$.}
	\label{fig:FloquetModesVskxp}
\end{figure}

Apart from simply agreeing with the past work of \cite{selvanayagam2016engineering}, we can easily extend the graphical/numerical investigation into new territories. Figure \ref{fig:FloquetModesVskxp} again shows a few $\mathrm{T}$ and $\Gamma$ coefficients versus incident angle at the design frequency (we now represent the incident angle with $k'_x/k$). Some interesting features can be observed. For example, at $k'_x/k = 0.53$ the curves exhibit a sudden change in behavior -- this occurs because at $k'_x/k = 0.53$ the first Floquet mode transitions from propagating to evanescent.

\begin{figure}[tbh!]
	\centering
	\includegraphics[width=0.6\textwidth]{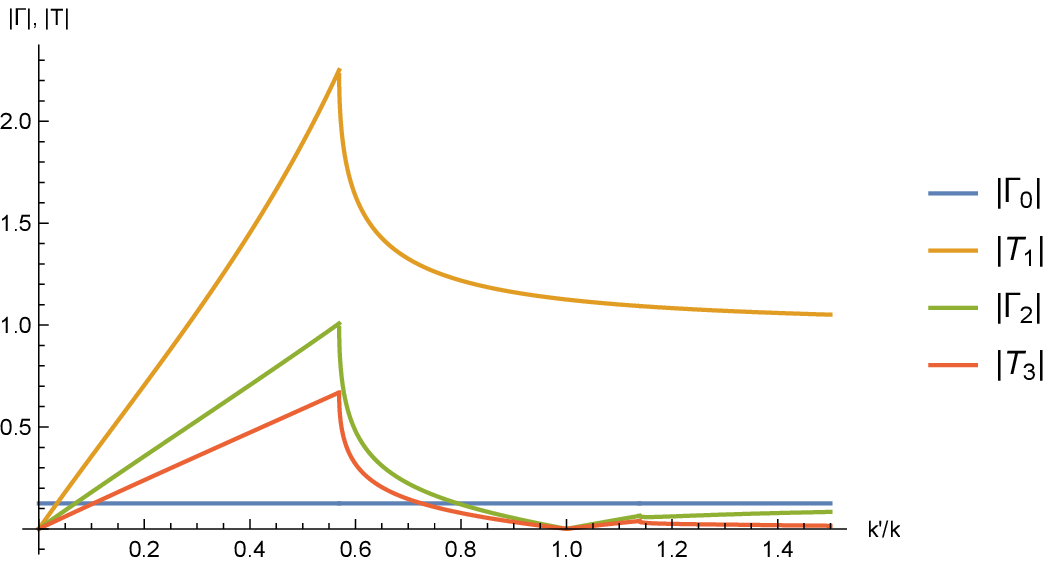}
	\caption{A few $|\mathrm{T}_a|$ and $|\Gamma_a|$ versus frequency $k'/k$ (at designed incidence angle) for a lossless surface with $k_s=0.47k$ and $k_x=0.174k$.}
	\label{fig:FloquetModesVskp}
\end{figure}

As previously mentioned, our analytical solution extends current understanding of tangent $\sigma\tau$ surfaces to scattering at any frequency. Figure \ref{fig:FloquetModesVskp} plots the same $|\mathrm{T}|$ and $|\Gamma|$, but versus frequency (at the designed angle of incidence). Indeed, our previous claim (found in the main text) that $\Gamma_0$ is frequency-independent is shown to be true graphically. 

\subsection{Proof of Field Non-Convergence for True Tangent Surfaces}

At this point all that is left is to rigorously prove that the scattered fields of a lossless tangent $\sigma\tau$ boundary do not converge, as we claimed previously with somewhat incomplete arguments. These non-convergence claims occur twice -- first when discussing designed incidence (see (\ref{eq:FieldConvDesInc}) and the discussion around) and then for non-designed incidence (see (\ref{eq:IdealRefFieldConvergence2}) and the discussion around). In both cases, although slightly different, the large-mode number amplitudes $E^{J/M}_{|a|\gg 1}$ exhibit similar behavior. And so, let us first consider the convergence of fields with Floquet amplitudes given by (\ref{eq:IdealRefFieldConvergence2}a):
\begin{align} \label{eq:Eagg1andFieldConv}
	E^J_{a\gg 1} &= A(x) \frac{(-1)^a}{k_a'+k_1}\prod_{n=b}^{a-1} \frac{k_n'-k_1}{k_n'+k_1},   &   \lim_{\substack{b \to \infty \\ b < \infty}}\sum_{a=b}^\infty E^J_a e^{-j(k'_x + k_s a) x},
\end{align}
where the left equation reiterates (\ref{eq:IdealRefFieldConvergence2}a) and the right expression is the convergence-defining part of the total field which we will eventually want to consider. Note that the right expression is evaluated at $y=0$ (on the surface) -- we will study the convergence of fields on the surfaces themselves. 

First, let us evaluate $\prod_{n=b}^{a} \frac{k_n'-k_1}{k_n'+k_1}$ (note that $a>b$). Since we will be interested in large $b$, we can approximate $k'_{n\gg 1}\approx -j k_s n$. Thus, the product becomes $\prod_{n=b}^{a} \frac{-j k_s n-k_1}{-j k_s n+k_1}$. Note that $k_1$ is real ($\sigma\tau$ surfaces are designed to achieve refraction between propagating waves). Let us now re-write the numerators and denominators appearing in the product as
\begin{equation}
	\prod_{n=b}^{a} \frac{-j k_s n-k_1}{-j k_s n+k_1} = \prod_{n=b}^{a} \frac{e^{j\left(\pi+\atan \frac{k_s n}{k_1}\right)}}{e^{-j\atan \frac{k_s n}{k_1}}} = \prod_{n=b}^a (-1)e^{j2\atan \frac{k_s n}{k_1}} = (-1)^{a-b+1} e^{j2 \sum_{n=b}^a \alpha_n}.
\end{equation}
where we have defined the partial angles $\alpha_n \equiv \atan\frac{k_s n}{k_1}$. Since $k_s n \ll k_1$ we can say that $\alpha_n \lesssim \pi/2$. And so, let us provide an approximation for $\alpha_n$ in this regime. We rewrite the definition as
\begin{equation}
	\tan\alpha_n=\frac{\sin\alpha_n}{\cos\alpha_n}=\frac{k_s n}{k_1}.
\end{equation}
Sine and cosine can be expanded via Taylor series (up to first order) at the operating point $\pi/2$ to give 
\begin{equation}
	\frac{1}{\pi/2-\alpha_n}=\frac{k_s n}{k_1} \hphantom{aaa}\to\hphantom{aaa}  \alpha_n=\frac{\pi}{2}-\frac{k_1}{k_s n}.
\end{equation}
Using the above approximation, we can say that the original product has the simple form
\begin{equation}
	\prod_{n=b}^{a} \frac{k_n'-k_1}{k_n'+k_1} = (-1)^{a-b+1} e^{j2 \sum_{n=b}^a \left(\frac{\pi}{2}-\frac{k_1}{k_s n}\right)} = e^{-j \frac{2k_1}{k_s}\sum_{n=b}^a \frac{1}{n}}.
\end{equation}
At this point it is clear that for any finite $a$ and $b$ (with $a>b$), the above product is a complex number with unit magnitude (i.e. $\left|\prod_{n=b}^{a} \frac{k_n'-k_1}{k_n'+k_1}\right|=1$). Taking the limit of $a \to \infty$, we can safely say
\begin{equation}
	\lim_{a\to\infty}\left| \prod_{n=b}^{a} \frac{k_n'-k_1}{k_n'+k_1} \right| = 1.
\end{equation}
This result was used (without proof at the time) in our crude convergence assessment that appears below (\ref{eq:FieldConvDesInc}). On a related note, let us mention the following. While we have argued that the limit (as $a\to\infty$) of the absolute value of the product exists, the same cannot be said of the product itself. This is so because in this limit the phase of the result diverges -- the phase, given by $-\frac{2k_1}{k_s}\sum_{n=b}^\infty \frac{1}{n}$, suffers from the non-converging sum $\sum_{n=b}^\infty \frac{1}{n}$. And so, the product has no phase, and thus cannot be placed on the complex plane. In other words,
\begin{equation}
	\lim_{a\to\infty} \prod_{n=b}^{a} \frac{k_n'-k_1}{k_n'+k_1}  = \text{Does Not Exist}.
\end{equation}

At this point we have everything required to evaluate field convergence (given by the right expression of (\ref{eq:Eagg1andFieldConv})). Plugging in the mode amplitudes (and making use of our recent results), we obtain
\begin{equation}
	\lim_{\substack{b \to \infty \\ b < \infty}}\sum_{a=b}^\infty \left(\frac{j A(x)e^{-jk'_x x} }{k_s} (-e^{ -j k_s x})^a\frac{e^{-j \frac{2k_1}{k_s}\sum_{n=b}^{a-1} \frac{1}{n}}}{a} \right).
\end{equation}
Let us evaluate the behavior of this sum at the particular points $x=\pi/k_s+2\pi n$, with $n$ being an integer. At these points $(-e^{ -j k_s x})=1$ and the sum takes the form (we also disregard the common to all terms $j A(x)e^{-jk'_x x} /k_s$)
\begin{equation}
	\lim_{\substack{b \to \infty \\ b < \infty}}\sum_{a=b}^\infty \left( \frac{e^{-j \frac{2k_1}{k_s}\sum_{n=b}^{a-1} \frac{1}{n}}}{a} \right).
\end{equation}
We now ask whether this limit exists. Note that originally, we had to invoke the complicated-looking limit of $b\to\infty$ simply to bring the original sum to its current form. Now that we simply ask about its convergence, the limit of $b$ becomes irrelevant -- if the fields do not converge with any finite $b$ then in the limit of large $b$ they will also not converge. Also, it is enough to show non-convergence with a single choice of finite $b$, since any other finite $b$ is obtained by adding finitely many terms (an operation which does not affect convergence). And so, we choose to investigate the sum
\begin{equation}
	\sum_{a=1}^\infty \frac{e^{-j \frac{2k_1}{k_s}\sum_{n=1}^{a-1}\frac{1}{n}}}{a}.
\end{equation}
Note that when $a=1$, we interpret $\sum_{n=1}^{0}\frac{1}{n}=0$. We define the partial sums of the above as
\begin{equation}
	S_p = \sum_{a=1}^p \frac{e^{-j \frac{2k_1}{k_s}\sum_{n=1}^{a-1}\frac{1}{n}}}{a}.
\end{equation}
For clarity, let us evaluate a few of the partial sums: $S_1 = 1$, $S_2 = 1+\frac{1}{2}e^{-j2k_1/k_s}$, $S_3 = 1+\frac{1}{2}e^{-j2k_1/k_s}+\frac{1}{3}e^{-\frac{j2k_1}{k_s}\left(1+\frac{1}{2}\right)}$ and so on (with a clear pattern emerging). Imagine plotting each $S_p$ in the complex plane -- the first point is at 1, then the second is a distance $1/2$ away (at an angle $-j2k_1/k_s$), the third yet another $1/3$ and so on. When $p$ is large, each successive increment moves the point a very small distance. Thus, for large $p$, $S_p$ resembles a continuous curve in the complex plane. Another important observation becomes apparent when we write the difference between neighboring partial sums:
\begin{equation}
	S_p-S_{p-1} = \frac{1}{p} e^{-j\frac{2k_1}{k_s}\sum_{n=1}^{p-1} \frac{1}{n}} = \frac{1}{p}e^{j\angle(S_{p-1}-S_{p-2})-j\frac{2k_1}{k_s(p-1)}},
\end{equation}
where we noted that each successive increment in the curve gains a small angle $\frac{2k_1}{k_s(p-1)}$ on the direction of the previous increment (hence the $\angle(S_{p-1}-S_{p-2})$). This difference $S_p-S_{p-1}$ can be considered as the differential of the ``continuous" curve. For aesthetic purposes we now refer to the curve as $c(l)$ (parameterized by a length along the curve $l$), and we say that at any point on the curve the direction of the differential is given by $\theta(l)$. With these new definitions, the above differential equation takes the form
\begin{equation}
	c(l +\d l)-c(l) = \d l \hphantom{.}e^{j\theta(l-\d l)-j\frac{2k_1 \d l}{k_s-\d l}},
\end{equation}
which we re-write as
\begin{equation}
	\frac{\d c}{\d l} = e^{j\theta(l-\d l)-j\frac{2k_1 \d l}{k_s-\d l}} = e^{j\theta(l)-j\frac{2k_1 \d l}{k_s}},
\end{equation}
where we omitted infinitesimal quantities in the final expression. We can also find the explicit form of $\theta(l)$ by writing out its difference equation:
\begin{equation}
	\theta(l+\d l) -\theta(l) = -\frac{2k_1 }{k_s}\d l \hphantom{aaa}\to\hphantom{aaa} \frac{\d \theta}{\d l}=-\frac{2k_1}{k_s} \hphantom{aaa}\to\hphantom{aaa} \theta(l) = -\frac{2k_1}{k_s} l + C,
\end{equation}
where $C$ is an integration constant (without loss of generality let us set $C=0$). Plugging in the explicit $\theta(l)$ into the differential equation for the curve we obtain
\begin{equation}
	\frac{\d c}{\d l} = e^{-j\frac{2k_1}{k_s}l},
\end{equation}
where we again omitted an infinitesimal term in the exponent. We now wish to know what kind of curves conform to this differential equation in the complex plane. To answer this, consider a circle (which we refer to as $c'(l)$) of radius $k_s/(2k_1)$ centered on the origin. If we parameterize this circle by length starting from the top (at $jk_s/(2k_1)$) in a clockwise fashion we obtain the following curve equation
\begin{equation}
	c'(l) = \frac{k_s}{2k_1} \left( \sin\frac{2k_1 l}{k_s} +j \cos\frac{2k_1 l}{k_s} \right) \hphantom{aaa}\to\hphantom{aaa} \frac{\d c'}{\d l} = e^{-j\frac{2k_1}{k_s}l}.
\end{equation}
As shown above, upon differentiation of $c'(l)$, we end up with our original differential equation -- showing that circles conform to our equation for $c(l)$. This shows that indeed, the fields of the lossless tangent $\sigma\tau$ surface do not converge as the Floquet modes are summed. Instead, for example at positions given by $x=\pi/k_s+2\pi n$ and $y=0$, the electric field phasor simply traces a circle in the complex plane as the modes are summed with no limiting value.

\begin{figure}[tbh]
	\centering
	\includegraphics[width=0.6\textwidth]{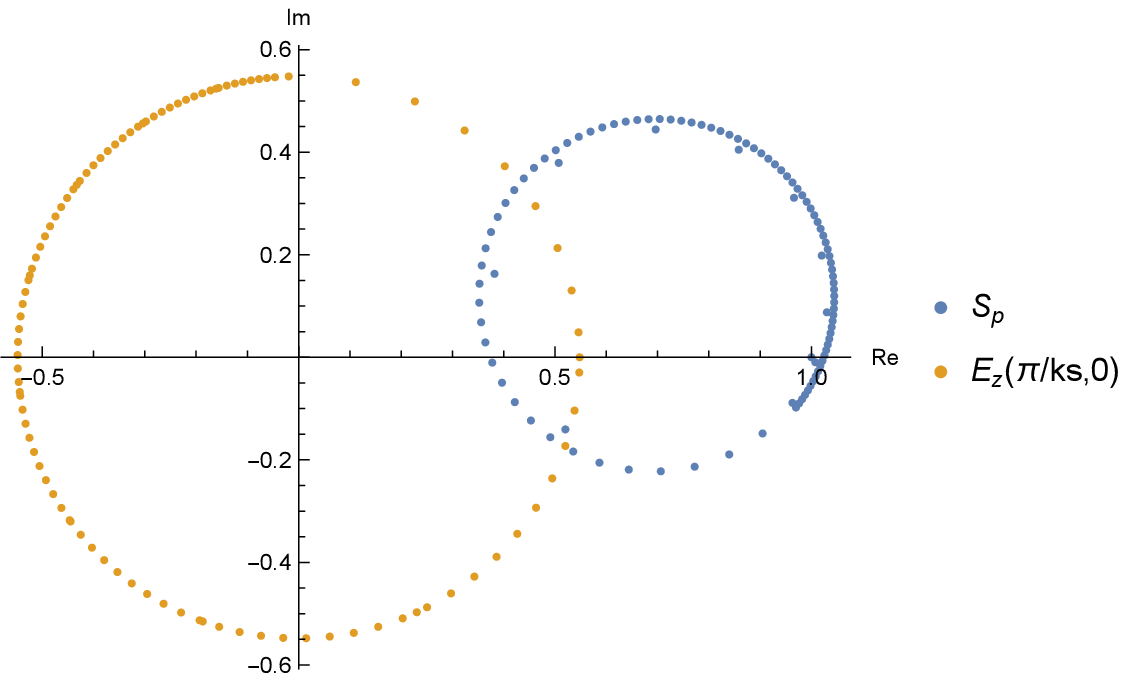}
	\caption{$S_p$ and partial field (up to $p^\text{th}$ Floquet mode) at $x=\pi/k_s$ and $y=0$, evaluated with $p$ ranging from 1 to 100. The surface and incidence for which the partial field is calculated are given by $k_s=0.5k$, $k_x=0.2k$, $k'=k$ and $k'_x=0.5k$.}
	\label{fig:FieldNonConvCircles}
\end{figure}

Let us remind the reader that we have performed the above convergence analysis for $E^J_{a \gg 1}$ modes. The convergence analysis for the lower half of the modes (i.e. $E^J_{a\ll -1}$) is the same, except some minor differences appear (for example the field phasors will trace counter-clockwise circles instead of the clockwise ones). And finally, let us show the proved non-convergence graphically. Consider a surface designed with $k_s=0.5$, $k_x=0.2k$ and illuminated with $k'=k$ and $k'_x=0.5k$. In Fig. \ref{fig:FieldNonConvCircles} we plot the points of $S_p$ and $E_i e^{-j k'_x x}+\sum_{a=0}^p E^J_a e^{-j (k'_x+k_s a) x}$ (i.e. the incident wave plus a partial sum of Floquet modes at the surface coordinate $x=\pi/k_s$) from $p=1$ to 100. Indeed, both the partial sum $S_p$ and the field on the surface do not converge, with a circular trajectory evident in the complex plane.

\bibliographystyle{IEEEtran}
\bibliography{PRLSupplementaryBib}